\documentclass[twocolumn]{aastex63}
\usepackage{enumitem}
\usepackage{amsfonts}
\usepackage{amsmath}
\usepackage{multirow}
\usepackage{url}
\usepackage{threeparttable}

\def\myerr[#1]{{\color{red} #1}}
\def\myemph[#1]{{\color{blue} #1}}
\def\hjmo[#1]{{\color{green} #1}}
\def\myrevise[#1]{#1}
\def\myrevises[#1]{#1}
\def\term[#1]{{\bf \ttfamily #1}}

\setlist[itemize,1]{label=$\bullet$}
\setlist[itemize,2]{label=$\bullet$}
\setlist[itemize,3]{label=$\bullet$}
\setlist[itemize,4]{label=$\bullet$}
\setlist[itemize]{leftmargin=*}
\def\bit{\begin{itemize}[topsep=0em,parsep=0em,itemsep=0em,partopsep=0em,leftmargin=2em]}
	\def\eit{\end{itemize}}

\def\beq{\begin{equation}}
\def\eeq{\end{equation}}
\def\bey{\begin{eqnarray}}
\def\eey{\end{eqnarray}}

\def\bfrm[#1]{\mathrm{{\bf#1}}}
\newcommand{\norm}[1]{\left\lVert#1\right\rVert}
\def\oderiv{{\rm d}}

\def\gs{\mathrel{\raise1.16pt\hbox{$>$}\kern-7.0pt
		\lower3.06pt\hbox{{$\scriptstyle \sim$}}}}
\def\ls{\mathrel{\raise1.16pt\hbox{$<$}\kern-7.0pt
		\lower3.06pt\hbox{{$\scriptstyle \sim$}}}}
\def\gtsima{\, {\buildrel > \over \sim} \,}
\def\ltsima{\, {\buildrel < \over \sim} \,}
\def\prosima{\, {\buildrel \propto \over \sim} \,}
\def\gsim{\lower.5ex\hbox{\gtsima}}
\def\lsim{\lower.5ex\hbox{\ltsima}}
\def\simgt{\lower.5ex\hbox{\gtsima}}
\def\simlt{\lower.5ex\hbox{\ltsima}}
\def\simpr{\lower.5ex\hbox{\prosima}}

\def\mpc{\, h^{-1}{\rm {Mpc}}}

\def\msun{\, h^{-1}{\rm M_\odot}}

\def\Rvir{R_{\rm vir}}
\def\Vvir{V_{\rm vir}}
\def\Mhalo{M_{\rm halo}}

\def\halospin{\lambda_{\rm s}}
\def\haloshape{q_{\rm axis}}
\def\halopc[#1]{{\rm PC}_{\rm MAH, #1}}

\shorttitle{Halo Properties}
\shortauthors{Chen et al.}
\graphicspath{{./}{figures/}}

\def\mahmz{M_{z}}
\def\mahsz{s(z)}
\def\mahsvec{\mathrm{\bf s}}
\def\mahpcvec{ \mathrm{\bf pc}_{\rm MAH} }

\begin{document}

\title{Relating the structure of dark matter halos to their assembly and environment}

\correspondingauthor{Yangyao Chen}
\email{yangyao-17@mails.tsinghua.edu.cn}

\author[0000-0002-4597-5798]{Yangyao Chen}
\affiliation{Department of Astronomy, Tsinghua University, Beijing 100084, China}
\affiliation{Department of Astronomy, University of Massachusetts, Amherst, MA 01003-9305, USA}

\author[0000-0001-5356-2419]{H.J. Mo}
\affiliation{Department of Astronomy, University of Massachusetts, Amherst, MA 01003-9305, USA}

\author[0000-0002-8711-8970]{Cheng Li}
\affiliation{Department of Astronomy, Tsinghua University, Beijing 100084, China}

\author[0000-0002-4911-6990]{Huiyuan Wang}
\affiliation{Key Laboratory for Research in Galaxies and Cosmology, Department of Astronomy, University of Science and
	Technology of China, Hefei, Anhui 230026, China}
\affiliation{School of Astronomy and Space Science, University of Science and Technology of China, Hefei, Anhui 230026, China}

\author[0000-0003-3997-4606]{Xiaohu Yang}
\affiliation{Department of Astronomy, School of Physics and Astronomy, Shanghai Jiao Tong University, Shanghai, 200240, China}
\affiliation{Tsung-Dao Lee Institute, and Shanghai Key Laboratory for Particle Physics and Cosmology, Shanghai Jiao Tong University, Shanghai, 200240, China}

\author[0000-0003-1967-4091]{Youcai Zhang}
\affiliation{Key Laboratory for Research in Galaxies and Cosmology, Shanghai Astronomical Observatory, Shanghai 200030, China}

\author[0000-0002-3775-0484]{Kai Wang}
\affiliation{Department of Astronomy, Tsinghua University, Beijing 100084, China}
\affiliation{Department of Astronomy, University of Massachusetts, Amherst, MA 01003-9305, USA}

\begin{abstract}
	We use a large $N$-body simulation to study the relation of the structural
	properties of dark matter halos to their assembly history and environment. 
	The complexity of individual halo assembly histories can be well described 
	by a small number of principal components (PCs), which, compared to 
	formation times, provide a more complete description of halo assembly histories  
	and have a stronger correlation with halo structural properties.
	Using decision trees built 
	with the random ensemble method, we find that about $60\%$, 
	$10\%$, and $20\%$ of the variances in halo concentration, axis 
	ratio, and spin, respectively, can be explained by combining four 
	dominating predictors: the first PC of the  
	assembly history, halo mass, and two environment parameters. 
	Halo concentration is dominated by halo assembly. 
	The local environment is found to be important for the axis ratio
	and spin but is degenerate with halo assembly. The small percentages 
	of the variance in the axis ratio and spin   
	that are explained by known assembly and environmental 
	factors suggest that the variance is produced by many 
	nuanced factors and should be modeled as such.  
	The relations between halo intrinsic properties and environment 
	are weak compared to their variances, with the anisotropy of the
	local tidal field having the strongest correlation with halo 
	properties. Our method of dimension reduction and regression can 
	help simplify the characterization of the halo population 
	and clarify the degeneracy among halo properties.
\end{abstract}

\section{Introduction}
\label{sec_intro}

In the concordant $\rm \Lambda$ cold dark matter ($\rm \Lambda$CDM) cosmology, 
dark matter halos, the dense clumps formed through gravitational collapse 
of the initial density perturbations, are the basic building blocks 
of the cosmic web. The formation history of a halo not only depends on 
the properties of the local density field, but may also be affected by 
the environment within which it forms. Since galaxies are believed to form in the 
gravitational potential wells of dark matter halos, 
the halo population provides a link between the dark  
and luminous sectors of the universe. Consequently,  
the understanding of the formation, structure, and environment of dark 
matter halos and their relations to each other has long been 
considered one of the most important parts of galaxy 
formation \citep[e.g.][]{MoBoschWhite:2010:GFE}.

Dark matter halos are diverse in their structure, mass assembly history (MAH), 
and interaction with the large-scale environment. Among the 
structural properties of dark matter halos, the most important ones 
are the concentration parameter~\citep{NavarroJ_FrenkC_WhiteS_1997_NFWProfile}, 
the spin parameter \citep{BettPhilip:2007:HaloSpinShapeInMillSim:BiasOnSpin,
	GaoL:2007:HaloAssemblyBiasOnSpinOrSub, MacCio2007}, and the shape parameter 
\citep{JingYP:SutoY:2002:ApJ:HaloDensityProfile:Shape,
	BettPhilip:2007:HaloSpinShapeInMillSim:BiasOnSpin,
	HahnOliver:2007:HaloSpinShapeMAHOnCosmicWebClassif,MacCio2007}. 
In $N$-body simulations, halo concentration is found to be correlated 
with halo mass and MAH~\citep{NavarroJ_FrenkC_WhiteS_1997_NFWProfile,
	JingYP:2000:HaloConcenScatterAndOnMAH,
	Wechsler:2002:HaloConcetrationAndMAHFitting,
	ZhaoDH:MoHJ:2003ApJ:NBodyValidationZMJBModel:HaloConcentration, 
	ZhaoDH:MoHJ:2003MN:ZMJBModel:HaloConcentration,
	MacCio2007,MacCioAV:2008:HaloStructVsCosmology, 
	ZhaoDonghai_2009_HaloConcenAnalytical, 
	LudlowAD:2014:HaloConcentrationAndHistory, 
	LudlowAD:2016:HaloConcentrationAndHistoryInWDM}. The spin and shape 
parameters are also found to be related to other properties, 
such as halo mass and large-scale
environment~\citep{MacCio2007,MacCioAV:2008:HaloStructVsCosmology,
	WangHuiyuan_MoHoujun_2011_HaloDependOnEnv}. However, these relations 
have large variance and remain poorly quantified. 

The mass assembly histories of halos in general are complex. 
In the literature, a common practice is to focus on the main-branch 
assembly histories, ignoring other branches
\citep[e.g.,][]{vandenBosch:2002:HaloUniversalMAH,
	Wechsler:2002:HaloConcetrationAndMAHFitting,
	ZhaoDH:MoHJ:2003ApJ:NBodyValidationZMJBModel:HaloConcentration, 
	ZhaoDH:MoHJ:2003MN:ZMJBModel:HaloConcentration,
	LudlowAD:2013:HaloProfAndHistory, 
	LudlowAD:2014:HaloConcentrationAndHistory, 
	LudlowAD:2016:HaloConcentrationAndHistoryInWDM}. 
Attempts have been made to describe the histories
of individual halos with simple parametric forms 
\citep{
	vandenBosch:2002:HaloUniversalMAH, 
	Wechsler:2002:HaloConcetrationAndMAHFitting,
	ZhaoDH:MoHJ:2003ApJ:NBodyValidationZMJBModel:HaloConcentration, 
	ZhaoDH:MoHJ:2003MN:ZMJBModel:HaloConcentration,
	Tasitsiomi:2004:HaloMAHAndProfileFitting, 
	McBrideJ:2009:HaloMAHFitting, CorreaCA:2015:OriginAndFittingHaloMAH,
	ZhaoDonghai_2009_HaloConcenAnalytical}. These simple models are useful in providing
some crude description of halo assembly histories, but are not meant 
to give a complete characterization. Because of this, 
a variety of formation times have also been defined to characterize 
different aspects of the assembly histories of dark matter halos
\cite[see, e.g.,][for a review]{LiYun_MoHoujun_2008_HaloFormationTimesDef}. 
These formation times, usually degenerate among themselves, again only 
provide an incomplete set of information about the full assembly history. 

The environment of a halo is also complex. To the 
lowest order, the average mass density around individual 
halos in a population can be used to characterize the 
distribution of the population relative to the underlying mass density field.
In general, the spatial distribution of halos (halo environment)
can depend on the intrinsic properties of the halos.  
The mass dependence, often called the halo bias
\citep{MoHJ:WhiteSDM:1996:HaloBias}, is a natural 
outcome of the formation of halos in a Gaussian density field
\citep{ShethRK:MoHJ:TormanG:2001:EllipseBias, 
	Zentner:2007:ExcursionSetTheory}. Furthermore, correlations
have also been found between halo bias and halo assembly history. 
Halos, particularly low-mass ones, that formed earlier tend to be more strongly 
clustered \citep{GaoL:SpringelV:2005:AssemblyBias, 
	GaoL:2007:HaloAssemblyBiasOnSpinOrSub, LiYun_MoHoujun_2008_HaloFormationTimesDef}.
This phenomenon is now referred to as the halo assembly bias, 
and various studies have been carried out to understand its 
origin \citep{
	SandvikHB:2007:OriginAssemblyBias, 
	WangHY:MoHJ:JingYP:2007:HaloEnv:HaloBias,
	WangHuiyuan:2009:EjectedSubhalosForAssemblyBias,
	Zentner:2007:ExcursionSetTheory, DalalN:2008:OriginAssemblyBias, 
	DesjacquesV:2008:EnvDepInEllipseCollapModel:OringinAssemblyBias, 
	LazeyrasT:2017:HaloAssemblyBiasOnJointTwo}. 
In addition to assembly history, halo bias has also been 
analyzed in its dependencies on other halo properties, such as halo concentration
\citep{WechslerRH:ZentnerAR:2006:AssemblyBiasAndOnConcenOrOccupation,
	JingYP:SutoY:MoHJ:2007:ApJ:ClusteringOnZform:Andc}, 
substructure occupation
\citep{WechslerRH:ZentnerAR:2006:AssemblyBiasAndOnConcenOrOccupation, 
	GaoL:2007:HaloAssemblyBiasOnSpinOrSub}, halo spin 
\citep{BettPhilip:2007:HaloSpinShapeInMillSim:BiasOnSpin, 
	GaoL:2007:HaloAssemblyBiasOnSpinOrSub, 
	HahnOliver:2007:HaloSpinShapeMAHOnCosmicWebClassif,
	WangHuiyuan_MoHoujun_2011_HaloDependOnEnv}, and 
halo shape~\citep{HahnOliver:2007:HaloSpinShapeMAHOnCosmicWebClassif,
	FaltenbacherA:2010:HaloAssemblyOnManyProp,
	WangHuiyuan_MoHoujun_2011_HaloDependOnEnv}. 
These dependencies are collectively referred to  
as the ``secondary bias", and sometimes also as the 
``assembly bias", presumably because these intrinsic properties 
may be related to halo formation.  

Since halo properties are intrinsically correlated, it is 
necessary to investigate the joint distribution of different 
properties. Along this line, \citet{JeasonDanielAkila:2011:PCA:HaloProperties} 
used rank-based correlation coefficients to quantify the correlation
between pairs of halo properties.
\citet{LazeyrasT:2017:HaloAssemblyBiasOnJointTwo} 
investigated halo bias as a function of two halo 
properties. They found that in all combinations of  
halo properties considered, halo bias can change with the 
second parameter when the first is fixed, and that the maximum of 
the halo bias occurs for halos with special combinations 
of the halo properties. For the environment, new parameters 
have also been introduced in addition to the local mass density. 
For example, \citet{WangHuiyuan_MoHoujun_2011_HaloDependOnEnv}
used local tidal fields to represent the halo environment, 
and \citet{SalcedoAN:2017:HaloNeighborBiasSpinBias} used the 
closest distance to a neighbor halo more massive than the halo
in question.
However, it is still unclear which quantity is the driving
force of halo bias and which combination of bias sources 
can provide a more complete representation. 
Answering these questions requires more advanced statistical 
tools to measure the capability of models to fit the data 
and to identify hidden degeneracy between model parameters.

Some statistical tools are available for such investigations.    
For unsupervised learning tasks, 
the {\sffamily Principal Component Analysis} (PCA) is a powerful tool 
to determine the main sources that contribute to the sample 
scatter and decompose the sample scatter along principal axes.
For example, \citet{WongAWC_2012_PCAHaloMAH} used this technique to reduce 
the dimension of halo MAH, while  
\citet{Cohn2015} and \citet{Cohn2018} applied the PCA to model the star formation 
history of galaxies. 
\citet{JeasonDanielAkila:2011:PCA:HaloProperties} used PCA to study 
the correlation of dark matter halo properties.
For supervised learning tasks, the 
{\sffamily Ensemble of Decision Trees} (EDT) or 
{\sffamily Random Forest} (RF) is capable of both classification
and regression. This method can also capture the nonlinear 
patterns, effectively reducing model complexity and 
discovering the dominant factors in target variables. 
RF has been widely used recently, for example, 
in identifying galaxy merger systems 
\citep{DelosRiosM:2016:SVM_RF_LR_MergerGalSystemIdentif}, 
in galaxy morphology classification 
\citep{Dobrycheva2017, SreejithS:2017:RF_SVM_NN_GalMorphologyClass},
in predicting neutral hydrogen contents of galaxies
\citep{RafieferantsoaM2017:SVM_RF_NN_boosting_PredictGalHI},
in determining structure formation in $N$-body simulations 
\citep[][see also \citet{LuisaLucieSmith:2019:BoostingOnDarkHaloFormation} for boosted trees]{LucieSmithL:2018:RF_DMParticleInNbody}, 
in measuring galaxy redshifts
\citep{StivaktakisR:2018:CNN_SpecRedshift}, 
in classifying star-forming versus quenched populations 
\citep{BluckAsaF_2019_MaNGA_QuenchingVSLocalOrGlobalEnv}, 
in identifying the best halo mass proxy in observation 
\citep{ManZhongyi:2019:RF_GalaxyHaloConnection}, 
and in estimating the star formation rate and stellar mass of galaxies 
\citep{BonjeanV:2019:RF_SFR_Mstar_Prediction}.

In this paper, we use both the PCA and the RF regressor to 
investigate the dependence of halo structural properties on 
halo assembly history and environment. The paper is organized as 
follows. In \S\ref{sec_haloquantities} we describe the simulation and the 
halo quantities to be analyzed. In \S\ref{sec_assembly} we demonstrate how 
to use PCA to extract information about halo assembly history.
In \S\ref{sec_struct_to_assembly} we relate
halo structure properties to assembly history and environment,  
identifying the dominating factors that determine halo properties. 
We also investigate the dependence of halo structure and 
assembly history on halo environment and halo mass. 
We summarize our main results in \S\ref{sec_summary}.

\section{Simulation and Halo Quantities} 
\label{sec_haloquantities}

\subsection{The Simulation} 
\label{ssec_simlation}

\begin{table*}
	\centering
	\caption{Samples Used in Our Analysis. The four columns are sample 
	identifier, number of halos, halo mass range, and usage, respectively. 
	The exact sample definitions can be found in \S\ref{ssec_simlation}. }
	\begin{tabular} { c|c|c|c } 
		\hline
		Sample & ${\rm N_{halo}}$ & $\Mhalo/[\msun]$ & Usage \\
		\hline\hline
		$\rm S_1$ & 2000 & $10^{[11, 11.2]}$ & \multirow{4}*{Samples with constrained halo masses. Used in \S\ref{ssec_assembly_PC}.}\\
		$\rm S_2$ & 1000 & $10^{[12, 12.2]}$& \\
		$\rm S_3$ & 500 & $10^{[13, 13.2]}$ &	\\
		$\rm S_4$ & 500 & $10^{[14, 14.5]}$ &	\\	
		\hline 
		$\rm S_c$ & 10000 & $ \geq 10^{11}$ & Mass-limited sample. Used in \S\ref{ssec_assembly_PC}. \\
		$\rm S_c'$ & 2335 & $\geq 5 \times 10^{11}$ & Subsample of $\rm S_c$, with all halo properties well-defined. Used in \S\ref{ssec_assembly_ftime}, \S\ref{ssec_struct_to_assembly}, \S\ref{ssec_struct_to_other}. \\
		$\rm S_L$ & 94524 & $10^{[11, 14.6]}$ & The 'larger' sample for binning statistics. Used in \S\ref{ssec_mass_on_structure}, \S\ref{ssec_env_assembly_bias}. \\
		\hline
	\end{tabular}
	\label{tab_def_samples}
\end{table*}

The $N$-body simulation used here is the 
ELUCID simulation carried out by 
\citet{WangHY2016_ConstrainedSim} using L-GADGET code, a memory-optimized 
version of GADGET-2 \citep{Springel2005}. The simulation uses
$3072^3$ dark matter particles, each with a mass $3.08\times10^8\msun$, 
in a periodic cubic box of $500$ comoving $\mpc$ on a side. 
The cosmology parameters used are those based on  
WMAP5 \citep{Dunkley2009}: a $\Lambda$CDM universe with density 
parameters $\Omega_{\rm K,0}=0$, $\Omega_{\rm M,0}=0.258$, 
$\Omega_{\rm B,0}=0.044$ and $\Omega_{\rm \Lambda, 0}=0.742$, 
a Hubble constant $H_0 = 100\ h\,{\rm km\,s^{-1}\,Mpc^{-1}}$ 
with $h=0.72$, and a Gaussian initial density field with power spectrum 
$P(k)\propto k^n$ with $n=0.96$ and an amplitude specified by 
$\sigma_8=0.80$. A total of 100 snapshots, uniformly spaced in  
$\log (1+z)$ between $z=18.4$ and $z=0$, are taken and stored.

Halos and subhalos with more than 20 particles are identified by the 
friends-of-friends \citep[FoF; see, e.g.,][]{DavisM1985_FOF} and SUBFIND 
\citep{Springel2005} algorithms. Halos and subhalos among different 
snapshots are linked to build halo merger trees.
\footnote{We use only halos and halo merger trees in this paper; subhalos are 
	not included in the samples.}
Halo virial radius $\Rvir$ is related to halo mass, $\Mhalo$, 
through 
\begin{equation}\label{eq_mhalo}
\Mhalo= {4\pi\Rvir^3\over 3} 
\Delta_{\rm vir} {\overline\rho}\,,   
\end{equation}
where ${\overline\rho}$ is the mean density of the universe, 
and $\Delta_{\rm vir}$ is an overdensity obtained from 
the spherical collapse model \citep{Bryan1998}.
The center of a halo is assumed to be the position of the most 
bound particle of the main subhalo. 
Halo mass $\Mhalo$ is computed by summing  
over all the particles enclosed within $\Rvir$. The virial velocity, 
$\Vvir$, is defined as $\Vvir=\sqrt{G\Mhalo/\Rvir}$, where $G$ 
is the gravitational constant. We use halos with masses 
$\ge 10^{10} \msun$ directly from the simulation, 
and we use {\sffamily Monte-Carlo}-based merger trees  
to extend the mass resolution of trees down to $10^9 \msun$
\cite[see][]{ChenYangyao_2019_CosmicVariance}.

From the halo merger tree catalog constructed above, we form 
four samples according to halo mass: 
${\rm S_1}$ contains 2000 halos with 
$10^{11} \le \Mhalo/\msun \le 10^{11.2}$; 
${\rm S_2}$ contains 1000 halos with 
$10^{12} \le \Mhalo/\msun \le 10^{12.2}$; 
${\rm S_3}$ contains 500 halos with 
$10^{13} \le \Mhalo/\msun \le 10^{13.2}$; and 
${\rm S_4}$ contains 500 halos with 
$10^{14} \le \Mhalo/\msun \le 10^{14.5}$.
We also construct a complete sample, ${\rm S_c}$,  
which contains 10,000 halos with $\Mhalo/\msun \ge 10^{11}$ 
to represent the total halo population.
\myrevise[All halos in samples $\rm S_1$, $\rm S_2$, $\rm S_3$, $\rm S_4$ and
$\rm S_c$ are randomly selected from simulated halos at $z=0$.] 
When halos need to be divided into subsamples   
according to some properties, the sample size of $\rm S_c$
may be insufficient. In this case, we construct a larger 
sample by the following steps.
Starting from all simulated halos at $z=0$ with 
$\Mhalo/\msun = 10^{11}$ - $10^{14.6}$, we divide 
them into mass bins of width of $0.3\,{\rm dex}$. 
If the number of halos in a bin exceeds 10,000, we randomly 
choose 10,000 from them; otherwise all halos in this mass bin 
are kept. This gives a sample of 94,524 halos, 
which is referred to as sample $\rm S_L$.
Due to the mass resolution of the ELUCID simulation, some properties 
of small halos cannot be derived reliably. Whenever these properties 
are needed, we use another halo sample, ${\rm S_c'}$, 
which contains all halos with $\Mhalo \ge 5\times10^{11}\msun$ 
in sample $\rm S_c$. 

Note that halos with recent major mergers may have structural properties 
that are very different from virialized halos, and including them in our sample 
will significantly increase the variance of halo properties, 
thereby affecting the statistics derived from the sample.
We use the criteria described in Appendix~\ref{sec_sim_unrelax} to 
exclude those 'unrelaxed' halos.

All of the samples used in this paper are summarized in the Table~\ref{tab_def_samples}.
Note that we use only a fraction of all halos in a given mass range 
available in the simulation to save computational time. 
We have made tests using larger samples to confirm that the 
samples we use are sufficiently large to obtain robust results.

\subsection{Halo assembly history} 
\label{ssec_assembly_def}

Following the literature~\citep[e.g.,][]{vandenBosch:2002:HaloUniversalMAH,
Wechsler:2002:HaloConcetrationAndMAHFitting, ZhaoDH:MoHJ:2003ApJ:NBodyValidationZMJBModel:HaloConcentration, 
ZhaoDH:MoHJ:2003MN:ZMJBModel:HaloConcentration}, we
define the halo MAH of a halo as the 
main-branch mass $\mahmz$ as a function of redshift $z$ in the halo 
merger tree rooted in that halo. Based on a theoretical consideration
of halo formation \citep[e.g.][]{ZhaoDonghai_2009_HaloConcenAnalytical}, 
we use the following quantity as the mass variable: 
\beq \label{eq_def_mahsz}
\mahsz=\sigma(M_{\rm 0})/\sigma(\mahmz),
\eeq
where $\sigma(M)$ is the rms of the $z=0$ linear 
density field at the mass scale $M$. 
Similarly, we use 
\beq \label{eq_def_deltac} 
\delta_{\rm c}(z)=\delta_{\rm c,0}/D(z),
\eeq
as the time variable, where $\delta_{\rm c,0}=1.686$ is the critical overdensity  
for spherical collapse, and $D(z)$ is the linear growth factor at $z$.
We use the transfer function given by \citet{Eisenstein1998} and the linear 
growth factor $D(z)$ from \cite{Carroll1992}.
These definitions for the mass and time variables are well motivated by 
the self-similar behavior of halo formation expected in 
the Press-Schechter formalism 
\citep[e.g.][]{PressSchechter:1974:PS,MoBoschWhite:2010:GFE}.

Thus, in our definition, the MAH of a halo is a vector 
$\mathrm{\bf s} = (s(z_1),\ s(z_2),\ ...,\ s(z_{\rm M}))^{\rm T}$, 
with each of its elements being the main-branch mass at a 
snapshot in the merger tree
	\footnote{To avoid confusion, 
	we use $\log$ to denote the base ten logarithm; 
	$\ln$ to denote the base $e$ logarithm; bold, roman lowercase characters 
	to denote vectors; bold, roman uppercase characters to denote matrix; 
	and $\norm{\cdot}$ to denote the 2-norm of a vector or a matrix.}. 
Such a high-dimensional vector is obviously too complex to be useful in 
characterizing the formation of a halo. To overcome this problem, 
a common practice is to characterize the full MAH by a set of formation times 
\citep[e.g.,][]{LiYun_MoHoujun_2008_HaloFormationTimesDef}. 
In our analysis, we will use both the formation times and 
the principal components (PCs; see \S\ref{sec_assembly}) to 
reduce the dimension of the MAH.

The MAH introduced above only uses the main branch of a halo merger tree, and 
thus it may potentially lose important information about the formation of a halo. 
However, our tests including the side branches (e.g., by modeling the 
assembly history with the progenitor mass distribution, as used in 
\citet{ParkinsonH2008_MCHaloTree}) showed that it does not provide 
important information about halo structural properties. 
We therefore only use the main-branch assembly history for our analysis.

\subsection{Halo concentration}
\label{ssec_concentration_def}

Dark matter halo profiles are usually modeled by a universal 
two-parameter form - the Navarro-Frenk-White (NFW) profile \citep{NavarroJ_FrenkC_WhiteS_1997_NFWProfile},
\beq \label{eq_def_nfw}
\rho_{\rm NFW}(r)= {{ \delta \rho_{\rm crit} }
\over {(r/r_{\rm s}) ( 1+ r/r_{\rm s})^2}},
\eeq
where $\rho_{\rm crit}=3H^2/(8\pi G)$ is the critical density of the universe.
This profile is specified by a dimensionless amplitude, $\delta$, and 
a scale radius, $r_{\rm s}$. The scale radius is usually expressed 
in terms of the virial radius, $\Rvir$, through a concentration 
parameter, $c\equiv \Rvir/r_{\rm s}$. Since both $\overline{\rho}$
and $\Delta_{\rm vir}$ in equation (\ref{eq_mhalo}) are known 
for a given cosmology, the profile of a halo can be specified 
by the parameter pair $(\Mhalo, \, c)$, where the halo mass $\Mhalo$ is defined 
in \ref{ssec_simlation}.

We obtain the concentration parameter of a halo through the following steps 
\citep[see][]{BhattacharyaSuman_HabibSalman_2013_HaloConcenMeasure}:
\begin{itemize}[leftmargin=*, itemsep=0pt, parsep=0pt, topsep=0pt]
	\item We divide the volume centered on a halo into $N_{\rm r}=20$ radial bins 
	equally spaced between 0 and $\Rvir$ (\myrevise[see \ref{ssec_simlation} for definitions of halo center and radius]), and calculate the mass 
	within each bin $i$, $M_{i}$, using the number of particles in the bin. 
	\item We compute the mass expected from the NFW profile in this bin, 
	$M_{i,{\rm NFW}}(c)$, assuming a concentration parameter $c$.
	\item We define an objective function, $\chi^2(c)$, to be minimized as
	\beq
	\chi^2(c)=\sum_{i=1}^{N_{\rm r}} \frac{[M_i-M_{i,{\rm NFW}}(c)]^2}{M_i^2/n_i}.
	\eeq
	\item We minimize the objective function to find the best-fit concentration 
	parameter $c_{\rm fit}={\rm argmin}_c\chi^2(c)$.  
\end{itemize}
In what follows, we will drop the subscript 'fit' and use $c$ to denote 
the concentration parameter obtained this way.  
\myrevise[As tested by \cite{BhattacharyaSuman_HabibSalman_2013_HaloConcenMeasure}, 
changing the radius range and binning scheme in the fitting 
only introduces a negligible difference in $c$. According to our tests, 
our results presented in the following sections are also 
insensitive to such changes.]

\subsection{Halo axis ratio}
\label{ssec_shape_def}

We model the mass distribution in a dark matter halo with an ellipsoid 
and use its axis ratio to characterize its shape. Our modeling 
consists of the following steps \cite[see][]{MacCio2007}:
\begin{itemize}[leftmargin=*, itemsep=0pt, parsep=0pt, topsep=0pt]
	\item We start from a FOF halo and calculate the spatial position 
	$\oderiv \bfrm[r]_i$ relative to the center of mass for each particle linked 
	to the halo. The inertia tensor $\mathcal{M}$ of the halo is given by the dyadic of 
	$\oderiv \bfrm[r]_i$ summing over all of the ${\rm N_p}$ particles in that halo:
	\beq
	\mathcal{M}=\sum_{i=1}^{\rm N_p} m_i\oderiv \bfrm[r]_i \oderiv \bfrm[r]_i^{\rm T},
	\eeq 
	where $m_i$ is the particle mass.
	\item We calculate the eigenvalues $\lambda_{\mathcal{M},i}$ and eigenvectors
	$v_{\mathcal{M},i}\,(i=1,2,3)$ of $\mathcal{M}$, and  
	rank the eigenvalues in descending order: $\lambda_{\mathcal{M},1}\ge 
	\lambda_{\mathcal{M},2} \ge \lambda_{\mathcal{M},3}$. So defined, 
	$v_{\mathcal{M},i}$ gives the axis direction of the inertia ellipsoid, and 
	$a_{\mathcal{M},i}$ = $\sqrt{\lambda_{\mathcal{M},i} }$ gives the length
	of the corresponding axis.
	\item We define the axis ratio $q_{\rm axis}$ of the halo as
	\beq
	q_{\rm axis}= \frac{a_{\mathcal{M},2}+a_{\mathcal{M},3}}{2a_{\mathcal{M},1}}.
	\eeq 
	So defined, $q_{\rm axis}=1$ for a spherical halo, and close to zero if the halo 
	is very elongated along the major axis.
\end{itemize}

\subsection{Halo spin}
\label{ssec_spin_def}

Following \citet{BullockJS_2001_HaloUniversalSpinProfile}, we define 
the spin parameter of a halo as
\beq 
\lambda_{\rm s} \equiv \frac{\norm{\bfrm[j]}}{ \sqrt{2}\Mhalo \Rvir \Vvir},
\eeq 
where $\Mhalo$, $\Rvir$ and $\Vvir$ are, respectively, the halo mass, virial radius, 
and virial velocity defined in \S\ref{ssec_simlation}. The 
total angular momentum $\bfrm[j]$ is defined as 
\beq
\bfrm[j] = \sum_{i=1}^{\rm N_p}m_i \oderiv \bfrm[r]_i \times \oderiv \bfrm[v]_i,
\eeq 
where $m_i$ is the particle mass, and $\oderiv \bfrm[r]_i$ and 
$\oderiv \bfrm[v]_i$ are particle position and 
velocity vectors relative to the center of mass, respectively.  
The summation is over all of the $\rm N_p$ particles linked in the FOF halo.

\subsection{Environmental quantities}
\label{ssec_environ_def}

Many definitions can be found in the literature to characterize the environment 
of a halo at different scales, traced by different objects, and including or excluding
the halo itself (see \citet{HaasMarcelR:2012:EnvironmentOnMass} for a review).
Here we define two quantities to describe the environment in which 
a halo resides: the density contrast as specified 
by the bias factor, and the tidal tensor. These quantities are computed 
directly from the N-body simulation. 

Firstly we follow~\citet{MoHJ:WhiteSDM:1996:HaloBias} to define 
halo bias $b$ as the ratio between the halo-matter cross-correlation function and the matter-matter autocorrelation function. For each
simulated halo $i$, we calculate its bias $b_{i}$ by
\begin{equation}
b_i = \frac{\xi_{{\rm hm}, i}(R)}{\xi_{\rm mm}(R)},
\end{equation}
\myrevise[where $\xi_{{\rm hm}, i}(R)$ is the overdensity 
centered at halo $i$ at a radius $R$, and $\xi_{{\rm mm}}(R)$ 
is the matter-matter auto-correlation function at the same radius
in the simulation at the redshift in question.
On linear scales, the bias factor is expected to depend 
only on halo mass, independent of $R$. 
We have checked the values of $b$ on 
different scales and found that 
the bias factor is almost constant at $R>5\mpc$.  
In the following, we compute both $\xi_{{\rm hm},i}$
and $\xi_{\rm mm}$ for $R$ between $5$ and $15\mpc$ at $z=0$, and 
we obtain the corresponding local linear bias $b_i$ for each halo 
using the above equation.]

For the tidal field, we follow \citet{RamakrishnanS:2019:CosmicWebAnisotropy:PrimaryBias} 
and define the halo environment in the following steps: 
we first divide the simulation box into a sufficiently fine 
grid (of $\rm N_{cell}^3$ grid points), and compute the density field 
$\rho(\bfrm[x])$ on each grid point using  
the clouds-in-cell method \citep{Hockney:1988:ParticleSimulation}. 
To describe the environment of a given halo, the density field is 
smoothed on some scale $R_{\rm sm}$ with a Gaussian kernel.
We then compute the potential field $\Phi(\bfrm[x])$ by solving the 
Poisson equation
\begin{equation}
\nabla^2\Phi(\bfrm[x])=4\pi G {\overline\rho}\delta(\bfrm[x]),
\end{equation}
where $\delta(\bfrm[x])=\rho(\bfrm[x])/\overline{\rho}-1$ is the overdensity 
and $\overline{\rho}$ the mean density of the universe.
Next we obtain the tidal field $\mathcal{T}(\bfrm[x])$ through
\begin{equation}
\mathcal{T}(\bfrm[x])=\nabla \nabla \Phi(\bfrm[x]).
\end{equation}
Finally, we solve the eigenvalue problem of the tidal tensor at 
each grid point to find the eigenvalues $\lambda_1(\bfrm[x])$, 
$\lambda_2(\bfrm[x])$, $\lambda_3(\bfrm[x])$ 
(ranked in a descending order). 

With all these, we obtain four environmental quantities for 
each halo: the local bias factor and the three eigenvalues 
of the tidal tensor: $\lambda_i$ ($i=1,2,3$). 
We follow the definition in~\cite{RamakrishnanS:2019:CosmicWebAnisotropy:PrimaryBias} to 
define the local tidal anisotropy at each halo's position as
\begin{equation}
\alpha_\mathcal{T}=\sqrt{q^2}/(1+\delta),
\end{equation}
where 
$q^2=\frac{1}{2}[(\lambda_1-\lambda_2)^2+(\lambda_2-\lambda_3)^2+(\lambda_1-\lambda_3)^2]$ 
is the halocentric tidal shear,
and $\delta=\lambda_1+\lambda_2+\lambda_3$ is the local overdensity.
For our applications, we choose $R_{\rm sm}=4R_{\rm vir}$, 
although we have checked that our 
conclusion does not change significantly by using other values of $R_{\rm sm}$.
We use $\rm N_{cell}=2560$, which is sufficiently fine for computing 
$\alpha_\mathcal{T}$ for the smallest halo in our sample. 

In Table~\ref{tab_def_haloprops}, we summarize the halo assembly, 
structure, and environmental properties used in our analysis. 

\begin{table*}
	\centering
	\caption{Summary of Halo Properties Studied in This paper. 
	Detail descriptions of halo MAH PCs and structural and environmental
	quantities can be found in \S\ref{sec_haloquantities} and \S\ref{ssec_assembly_PC}.}
	\begin{tabular} { c|c|c } 
		\hline
		Property Type & Notation & Meaning \\
		\hline\hline
		\multirow{1}*{Assembly History}
		& ${\rm PC}_{{\rm MAH},i}$ & $i$th principal component of MAH \\
		\hline
		\multirow{3}*{Structure} 
		& $c$ & concentration parameter of the NFW profile\\
		& $\lambda_{\rm s}$ & dimensionless spin parameter \\
		& $q_{\rm axis}$ & axes ratio of inertia momentum \\
		\hline
		\multirow{2}*{Environment} 
		& $b$ & halo bias factor \\
		& $\alpha_\mathcal{T}$ & tidal anisotropy \\
		\hline
	\end{tabular}
	\label{tab_def_haloprops}
\end{table*}

\section{Characterizing Halo Assembly History with PCs} 
\label{sec_assembly}

Because many parameters can be defined to characterize various 
aspects of the halo assembly history (e.g., different formation times), 
it is not feasible to use all of them directly to study the relationship 
between halo assembly and other halo properties. In this section, we 
describe a method that can be used to effectively reduce the 
dimension of the halo assembly history, which has some advantages over 
using formation times. 

We use the PCA described in Appendix \ref{ssec_method_pca} 
to reduce the dimension of the MAH. In this section we will show 
the advantages of using such a method in the following two aspects.  
(1) The PCs capture the most variance 
among all linear, low-dimension representations 
and are the best in reducing the reconstruction error. 
We will demonstrate their power in representing the halo MAH. 
(2) The PCs are tightly correlated with some formation times
widely used in the literature, which helps clarify the 
types of information PCs contain.
In the next section (\S\ref{sec_struct_to_assembly}), we will 
show that PCs are strongly correlated with halo structural 
properties, which helps us understand the origin of such properties. 
All these indicate that PCs are not only mathematically optimized 
approximations to the MAH but also have clear physical meanings.

\subsection{PCs of Halo Assembly Histories} 
\label{ssec_assembly_PC}

\begin{figure}
	\centering
	\includegraphics[width=7.5cm]{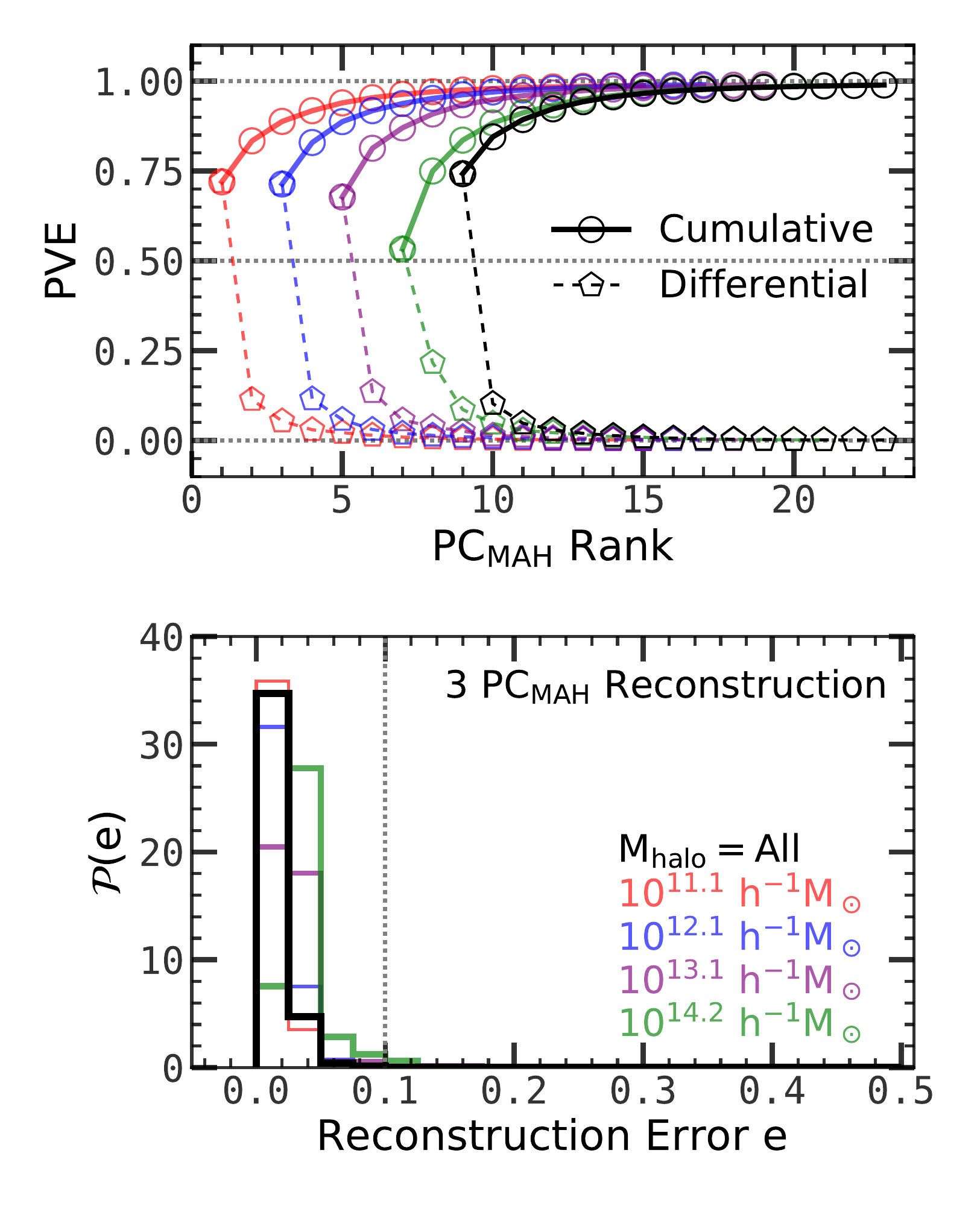}
	\caption{The performance of PCA on a halo MAH. {\bf Top:} PVEs and their cumulative version (see Appendix~\ref{ssec_method_pca} for definitions). Each curve is horizontally offset increasingly by 2 for clarity. {\bf Bottom:} distribution of reconstruction error $e$ using the first three MAH PCs. In both panels, five samples, $\rm S_1$, $\rm S_2$, $\rm S_3$, $\rm S_4$, $\rm S_c$, with different halo mass selections (see \S\ref{ssec_simlation} and Table~\ref{tab_def_samples}) indicated in the lower panel, are presented. }
	\label{fig:mah_pca_error}
\end{figure}

\begin{figure*}
	\centering
	\includegraphics[width=\textwidth]{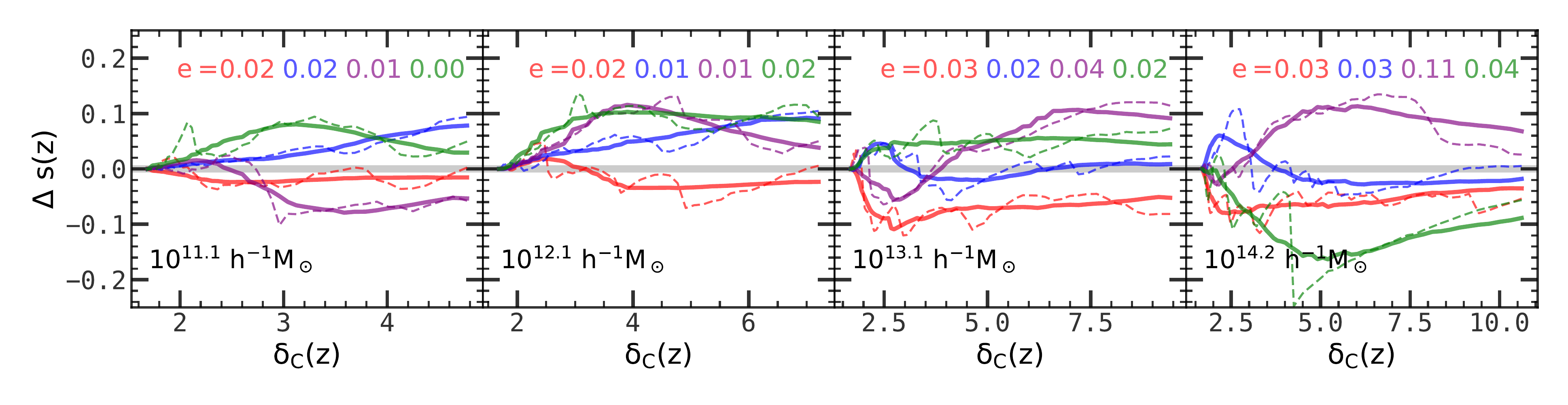}
	\caption{Reconstructed MAHs of example halos ({\bf solid}) compared with the real ones ({\bf dashed}). Four panels present halos with different masses indicated in each panel. In each case, the MAHs are reconstructed from the first three PCs. The reconstruction error $e$ for each halo is also presented. }
	\label{fig:mah_pca_recon_curve}
\end{figure*}

For a given sample of $N$ halos, the MAHs are represented by 
$\mathrm{\bf S} = (\mahsvec_1,\ \mahsvec_2,\ ...,\ \mahsvec_{\rm N})^{\rm T}$, 
where $\mahsvec_i$ is the MAH of the $i$th halo, as defined in 
\S\ref{ssec_assembly_def}. In simulation, the MAH of a halo 
is not traced below the mass resolution of the simulation. 
Halos of different mass therefore may be traced down to different 
snapshots, resulting in different lengths of the vector 
$\mahsvec_i$. For a given sample, we choose a snapshot where $90\%$ 
of the MAHs can be traced back to this time. MAHs extending beyond this snapshot 
are truncated, and MAHs that are terminated before this snapshot 
are padded with $0$. In this way, all of the $\mahsvec_i$ will have the 
same length, $M$, suitable for PCA. After applying the PCA to $\mathrm{\bf S}$, 
the $\mahsvec$ for each halo is transformed into a new coordinate system, 
producing a new vector that consists of a series of PCs, 
$\mahpcvec=(\halopc[1],\,\halopc[2],\,...,\,\halopc[M])$. 
In terms of the capability of capturing sample variance and reducing 
reconstruction errors, PCA is theoretically the optimal linear 
method. Thus, if the assembly histories of halos have 
some dominating modes, they are expected to be captured by the PCA.

The upper panel of Figure~\ref{fig:mah_pca_error} shows the PVE and CPVE curves 
(see Appendix~\ref{ssec_method_pca} for definitions) for the PCs of the 
five samples, ${\rm S_1}$, ${\rm S_2}$, ${\rm S_3}$, ${\rm S_4}$ and ${\rm S_c}$, 
defined in \S\ref{ssec_simlation} (see also Table~\ref{tab_def_samples}). 
Since the proportional explained variance (PVE) measures the fraction of sample variance explained by a 
PC, it is clear that the first several PCs, among all 
cases, can capture most of the variance in the halo MAH 
($>80\%$ by using the first three PCs). This demonstrates that strong 
degeneracy exists in the MAH of individual halos, suggesting that 
the assembly history can be described by using only a few eigen-modes. 
As shown by the CPVE curve, using a single parameter is insufficient 
to describe the MAH. It can at most explain as much variance as ${\rm PC}_1$, 
which is $55\%$ for the most massive halos ($\approx 10^{14}\msun$), 
and $70\%$ for the smallest halos ($\approx 10^{11}\msun$). 
In principle, we can add more PCs, which typically leads to  
better capture of the subtle structures in the MAHs.
The lower panel of Figure~\ref{fig:mah_pca_error} shows the distribution of 
the error $e$ in each sample when MAHs are reconstructed with
the first three PCs (see Appendix~\ref{ssec_method_pca} for the 
reconstruction algorithm). 
The reconstruction errors are almost all below $10\%$, demonstrating that 
the MAHs of halos can be represented well by a small 
number of PCs.

Figure~\ref{fig:mah_pca_recon_curve} shows some examples of 
the reconstructed MAHs using the first three PCs. Here 
$\Delta \mahsz=\mahsz-\overline{s}(z)$ is shown for several halos in 
each halo-mass-constrained sample (see \S\ref{ssec_simlation} and 
Table~\ref{tab_def_samples}), where $\overline{s}(z)$ is the mean MAH in 
the sample. The overall shape of the MAH is well captured by the reconstruction, 
although the MAHs of individual halos are quite diverse. 
Some fine structures in the MAH, caused by violent changes 
in the formation history due to merger events, are missed in the 
reconstruction. They can, in principle, be captured by including more PCs.

The rapidly converged PVE, the sharply peaked distribution of the 
reconstruction, and the well-reconstructed MAHs of individual halos 
all indicate that PCs are effective in reducing the dimension
of the halo MAH. In the following subsection, 
we will show the relation between PCs and some widely used 
halo formation times to gain more physical insights
into different PCs.

\subsection{PCs versus Halo Formation Times} 
\label{ssec_assembly_ftime}

The diversity of the assembly history shown in the 
Figure~\ref{fig:mah_pca_recon_curve} indicates that no single 
parameter can provide a complete description of the MAH.  
To reflect different aspects of the assembly history, 
different assembly indicators, for example, formation times, 
have been defined in the literature. These formation times 
are physically more intuitive compared with the more abstract PCs, 
although each of them only provides partial information 
about the MAH. Here we examine the relations between 
PCs and a number of halo formation times  
to gain some physical understanding of the PCs we obtain.

\begin{figure}
	\centering
	\includegraphics[width=\columnwidth]{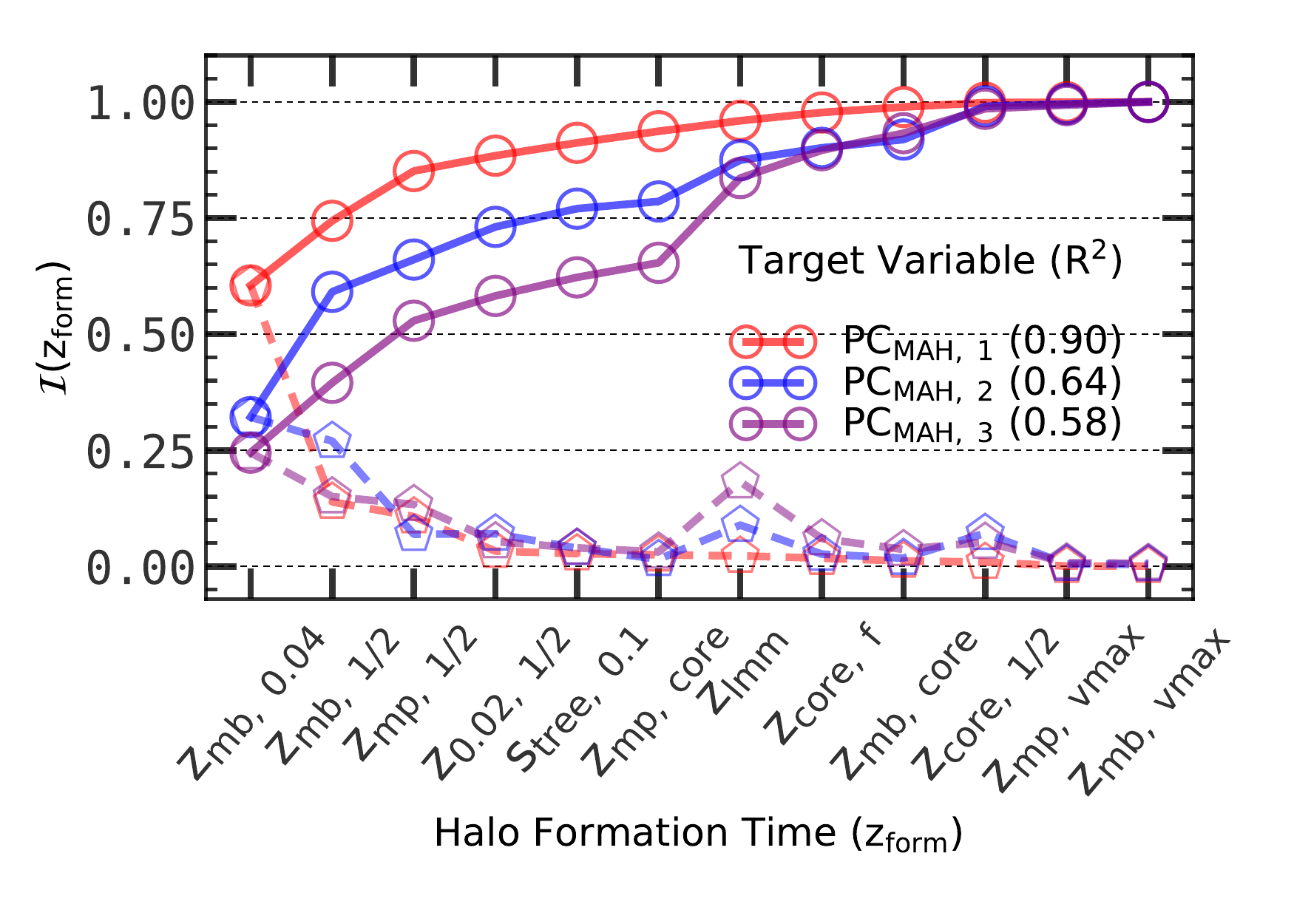}
	\caption{Performances ($R^2$) and contributions ($\mathcal{I}$) using formation 
	times to predict the first three ${\rm PC}_{\rm MAH}$s based on RF 
	regressors (see Appendix~\ref{ssec_method_rf} for the RF model). 
	{\bf Dashed}: importance values $\mathcal{I}$ of predictors. {\bf Solid}: the 
	cumulative of $\mathcal{I}$. The overall performances, $R^2$, are shown in 
	the parentheses in the legend. Halos are taken from the mass-limited
	sample $\rm S_c'$ with $\Mhalo \ge 5\times 10^{11} \msun $ (see \S\ref{ssec_simlation} and Table~\ref{tab_def_samples}). }
	\label{fig:mah_pca_ftime2pc_impcurve}
\end{figure}

In Appendix~\ref{sec_ftime_def} we list the formation times we use 
and their detailed definitions. 
Because of the large number of formation times 
and the potential nonlinear pattern in their relations 
with PCs, RF is an ideal tool for this task. 
In Appendix~\ref{ssec_method_rf} we describe the RF 
algorithm in detail. The two important outputs from the RF 
analysis are (1) the fraction of explained variance, $R^2$, 
and (2) the feature importance, $\mathcal{I}(x)$, for any predictor
variable $x$. These two quantities measure the performance of 
the regression and the contribution from each predictor variable 
in explaining the target variable, respectively. 
Figure~\ref{fig:mah_pca_ftime2pc_impcurve} shows how different formation 
times contribute to the diversity of the halo MAH. 
Here we use sample $\rm S_c'$ in which all of the formation times 
are well defined, as described in the 
\S\ref{ssec_simlation} (see also Table~\ref{tab_def_samples}), 
to regress the first three MAH PCs on all of the formation 
times. The performance, $R^2$, and the contribution $\mathcal{I}(x)$ of 
each formation time $x$ to each PC are plotted. For all PCs, 
the contribution from $z_{{\rm mb},0.04}$ is the most dominant, 
while $z_{{\rm mb},1/2}$ is also significant. 
However, the importance
of both decreases in higher order PCs.
Interestingly, the last major merger 
redshift, $z_{\rm lmm}$, which contributes little to 
${\rm PC}_{\rm MAH,1}$, is increasingly important as the PC order 
increases. For ${\rm PC}_{\rm MAH,3}$, the contribution from 
$z_{\rm lmm}$ is comparable to those from the other two formation times, 
$z_{{\rm mb},0.04}$, and $z_{{\rm mb},1/2}$. Since mathematically higher order 
PCs are capable of capturing more subtle patterns 
in the feature space, this behavior of the importance curves means that
assembly variables, such as $z_{{\rm mb},0.04}$, and $z_{{\rm mb},1/2}$,
mainly describe the low-order, overall patterns of the halo assembly history,
while major mergers are an important factor in producing the fine 
structure in MAHs.

\begin{figure}
	\centering
	\includegraphics[width=\columnwidth]{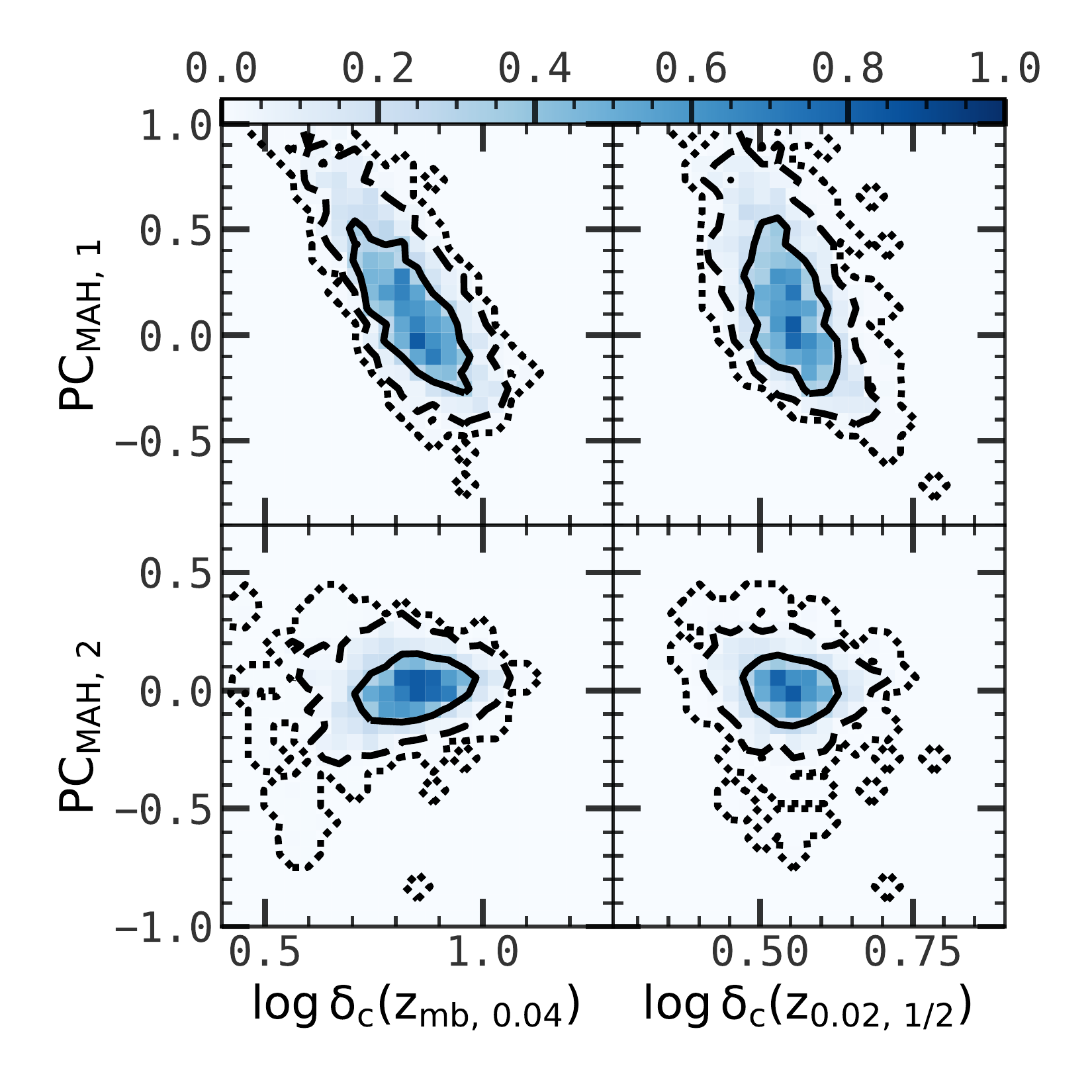}
	\caption{Relation of halo formation time $z_{{\rm mb},\ 0.04}$ ({\bf left}) or 
	$z_{{\rm 0.02},\ 1/2}$ ({\bf right}) to the first two MAH PCs. 
	Solid, dashed, and dotted contours cover 1, 2, 3 $\sigma$ regions, respectively. 
	Normalized 2D histograms are color-coded according to the color bar. Halos are 
	taken from the mass-limited sample $\rm S_c'$ with 
	$\Mhalo \ge 5\times 10^{11} \msun $ (see \S\ref{ssec_simlation} and 
	Table~\ref{tab_def_samples}). }
	\label{fig:mah_pca_ftime2pc_contour}
\end{figure}

The contribution curves in Figure~\ref{fig:mah_pca_ftime2pc_impcurve}
automatically suppress variable competition where multiple
degenerate feature variables compete in the prediction contribution
to the target variable. In the RF, if two variables compete but one of them
is slightly better, then the split algorithm prefers the better one and
gives it a higher importance value $\mathcal{I}(x)$. To show this more 
clearly, Figure~\ref{fig:mah_pca_ftime2pc_contour} plots 
the correlations between $z_{{\rm mb},0.04}$, $z_{0.02, 1/2}$ 
and the first two MAH PCs. Strong and nearly linear correlation 
between $\halopc[1]$ and $z_{{\rm mb}, 0.04}$, and 
between ${\rm PC}_{{\rm MAH},1}$ and $z_{0.02, 1/2}$ are seen, 
indicating that the variances in 
both two formation times contribute significantly to the variance in 
the MAH of the halos. Inspecting the contours, one can also see that 
the correlation between $\halopc[1]$ and $z_{{\rm mb}, 0.04}$ 
appears stronger. The larger contribution value to ${\rm PC}_{{\rm MAH},1}$ 
from $z_{{\rm mb}, 0.04}$ than from $z_{0.02,1/2}$ shown 
in Figure~\ref{fig:mah_pca_ftime2pc_impcurve} validates the 
strength of the correlation.

Figure~\ref{fig:mah_pca_ftime2pc_contour} also 
demonstrates that the description provided by a single  
halo formation time is incomplete. 
The large scatter seen in the contours of $\rm PC_{MAH,1}$ versus 
formation times means that the direction of the largest scatter 
in the MAH space is not fully aligned with the scatter caused by 
formation time, and that other parameters must also 
contribute to the distribution of halos in the MAH space. 
Moreover, compared to ${\rm PC}_{{\rm MAH}, 1}$, 
${\rm PC}_{{\rm MAH}, 2}$ has much weaker correlation 
with the halo formation times. Thus, the information 
contained in the second PC of MAH, which contributes
$10\%$-$20\%$ to the total variance as seen from 
the upper panel of Figure~\ref{fig:mah_pca_error}, is 
almost entirely missed when a single formation time is used 
to predict the MAH. 

The degeneracy and completeness problems can, in principle, 
be overcome if we use PCs to describe the assembly history, 
as PCs are linearly independent of each other. Typically, the 
use of a small number of low-order PCs can solve most of the 
problems associated with the regression of halo MAHs. 
If more subtle information is needed, one can always add more PCs 
without introducing too much degeneracy into the problem. 

\section{Relating Halo Structure to Assembly History and Environment} 
\label{sec_struct_to_assembly}

In this section, we investigate how halo structural properties 
are related to halo assembly history and environment. First, we
will examine which of the assembly history indicators correlates 
the best with halo structural properties. Second, we will 
show to what extent the halo structure
can be explained by assembly history and environment, 
and we answer the question whether there is a single dominating 
predictor for halo structure or if the predictors are degenerate in 
predicting halo structure. Third, we will show that our 
conclusion is valid even when halo mass is fixed. 
Finally, we revisit the problem of assembly bias, 
aiming to identify environment-assembly pairs of strong 
correlation.

\subsection{Halo Structure versus Assembly History} 
\label{ssec_struct_to_assembly}

It is well known that halo concentration is correlated with 
halo MAH
\citep[see e.g.,][]{NavarroJ_FrenkC_WhiteS_1997_NFWProfile, 
JingYP:2000:HaloConcenScatterAndOnMAH, 
Wechsler:2002:HaloConcetrationAndMAHFitting, 
MacCioAV:2008:HaloStructVsCosmology,
ZhaoDH:MoHJ:2003ApJ:NBodyValidationZMJBModel:HaloConcentration, 
ZhaoDH:MoHJ:2003MN:ZMJBModel:HaloConcentration, ZhaoDonghai_2009_HaloConcenAnalytical}. 
However, it is still unclear which single assembly parameter best predicts the 
concentration, and whether combinations of multiple parameters can improve 
the prediction precision. The same problem exists
when we consider other halo structural properties, for example, the axis ratio
$q_{\rm axis}$ and the spin parameter $\lambda_{\rm s}$. 

The difficulties involved here arise from the high dimension
of the feature space, the degeneracy or correlation among predictors, 
a possible nonlinear effect from predictors to target, 
and the ``bias-variance" trade-off in choosing model complexity. 
Again, {\sffamily Random Forest} can be used to tackle these 
problems (see Appendix~\ref{ssec_method_rf}). 
To this end, we build the regressor $y={\rm RF}(\bfrm[x])$, 
where $y$ is one of the three structural properties: 
$c$, $\haloshape$ or $\halospin$, 
and $\bfrm[x]$ are halo assembly indicators, 
either the first 10 MAH PCs, or all formation times. 
We also include halo mass in the predictor
variables, because it is treated as one of 
the major parameters in many halo-related problems. 
All these regressors are built based on sample $\rm S_c'$ 
(see \S\ref{ssec_simlation} and Table~\ref{tab_def_samples}) 
in which all formation times are well defined for all halos.

\begin{figure*}
	\centering
	\includegraphics[width=\textwidth]{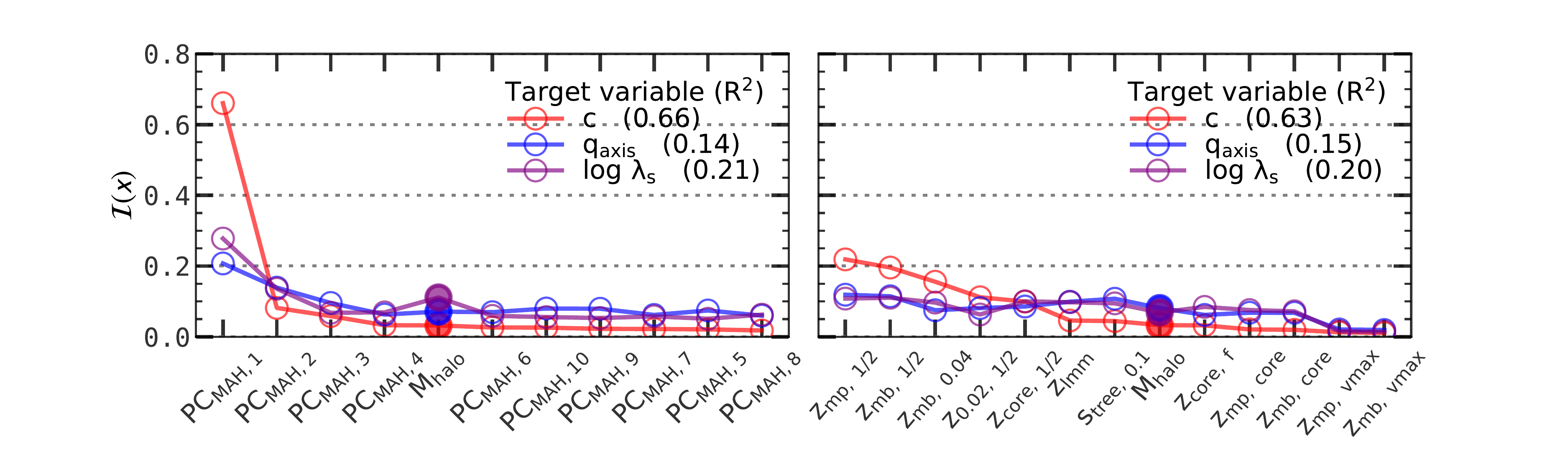}
	\caption{ Contributions $\mathcal{I}(x)$ in regressing halo concentration $c$ 
	({\bf red}), shape parameter $q_{\rm axis}$ ({\bf blue}), or spin parameter 
	$\log \lambda_{\rm s}$ ({\bf purple}) on halo assembly history 
	variables: the first 10 
	PCs of MAH ({\bf left} panel) or formation times ({\bf right} panel). 
	Halo mass is also included as a predictor variable and is represented 
	by filled symbols.
	For each case, $R^2$ indicates the overall performance of the regression. Halos are 
	taken from the mass-limited sample $\rm S_c'$ with $\Mhalo \ge 5\times 10^{11}\msun$ 
	(see \S\ref{ssec_simlation} and Table~\ref{tab_def_samples}).
	}
	\label{fig:struct_vs_assembly_impcurve}
\end{figure*}

The outputs of RF regressions, including the performance, $R^2$, 
and the contribution $\mathcal{I}(x)$ from each predictor variable, are shown in 
Figure~\ref{fig:struct_vs_assembly_impcurve}. 
As one can see from the red curves, 
when a large number of predictors are used, the upper 
limit in the prediction of $c$ using assembly history is about $65\%$. 
This indicates that the concentration parameter $c$ of a halo is 
largely determined by its assembly history. 
About $35\%$ of the variance is still missing if one uses only 
the mean relation to predict $c$ from assembly history. 
Furthermore, as seen from the left panel, the first MAH PC 
is by far the most important, accounting for about $67\%$ of the
total information provided by the MAH.
As a comparison, in the right panel, the combination of the 
three formation times, 
$z_{{\rm mb},1/2}$, $z_{{\rm mp},1/2}$ and $z_{{\rm mb},0.04}$, 
contains about the same amount of information as ${\rm PC}_{\rm MAH,1}$. 
Thus, if a single parameter is to be adopted as the predictor of $c$, 
${\rm PC}_{\rm MAH,1}$ is the preferred choice. 
We have also tried to combine halo formation times and PCs of the MAH as 
predictors, and we found that the overall performance changes little, 
indicating that the MAH PCs dominate the information content about 
halo concentration. 

For $q_{\rm axis}$ and $\lambda_{\rm s}$,
the upper-limit performances $R^2$, achieved by either PCs 
or formation times, are about $15\%$ and $20\%$, respectively,
much worse than $c$. No single variable seems to dominate 
the contribution, as indicated 
by the long and low tail in the $\mathcal{I}(x)$ plot. 
These suggest that the axis ratio and spin can be affected by many 
factors related to halo mass assembly, but the effects are 
all small. The similarity in behavior between $\lambda_{\rm s}$ 
and $q_{\rm axis}$ suggests that these two quantities may 
share parts of their origins. Indeed, 
as we will show below (\S\ref{ssec_struct_to_other}), 
$\lambda_{\rm s}$ and $q_{\rm axis}$ are strongly correlated, 
and both show strong correlation with the anisotropy of the 
local tidal field.

\begin{figure*}
	\centering
	\includegraphics[width=14cm]{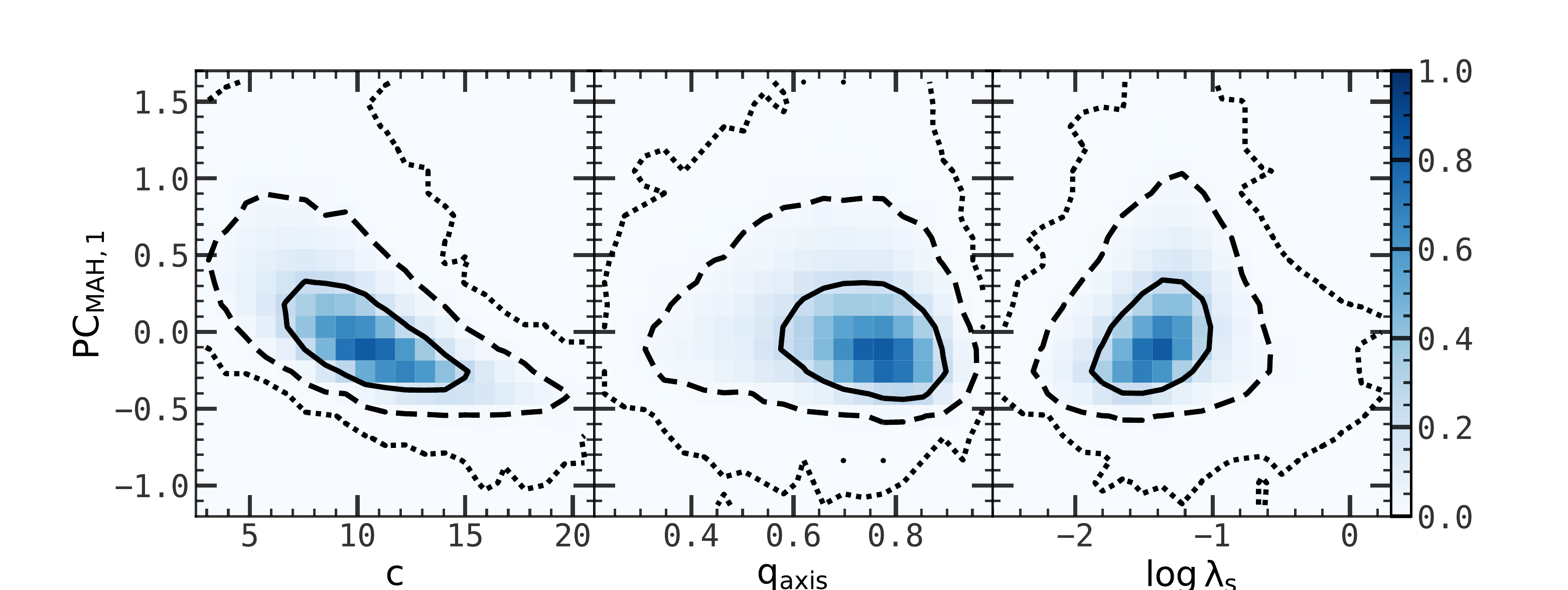}
	\caption{Relation of halo concentration $c$ ({\bf left}), shape parameter $q_{\rm axis}$ ({\bf central}), or the spin parameter $\log \lambda_{\rm s}$ ({\bf right}) and the first PC of the halo MAH. In each panel, the solid, dashed, and dotted contours cover the 1, 2, and 3$\sigma$ regions, respectively. The normalized 2D histograms are color-coded according to the color bar. Halos are taken from the mass-limited sample $\rm S_c'$ with $\Mhalo \ge 5\times 10^{11}\msun$ (see \S\ref{ssec_simlation} and Table~\ref{tab_def_samples}). }
	\label{fig:struct_vs_assembly_contour}
\end{figure*}

Figure~\ref{fig:struct_vs_assembly_contour} shows the distribution of halos
in the $\rm PC_{MAH,1}$ - structural parameter space. 
There is a strong trend that halos 
assembled late (large ${\rm PC}_{\rm MAH,1}$) tend to be less concentrated, 
and a weak trend that such halos tend to be more elongated 
and spin faster. These are all consistent with the output from 
the RF regressors and verify that the small contributions from assembly 
indicators to $\lambda_{\rm s}$ and $q_{\rm axis}$ are produced 
by the large scatter, rather than by variable competitions. 

\subsection{Environmental Effect and Spin-Shape Interaction} 
\label{ssec_struct_to_other}

We now add environmental predictors to the regression of the structural 
properties. 
To quantify the effects of variable competition, 
we adopt a commonly used approach called ``growing", where a series of 
regressors is built up with increasing number of predictors. 
(The approach is called ``pruning'', if the series runs reversely).
Whenever there is a tight correlation between an added predictor 
and the predictors already used, competition will show up 
as changes in their importance values, $\mathcal{I}(x)$, 
but the overall performance, $R^2$, will not be changed significantly
by the addition. 

As demonstrated in \S\ref{ssec_struct_to_assembly}, 
among all halo assembly indicators, the first PC of the halo MAH
is the dominating indicator for $c$. Even for $q_{\rm axis}$ and 
$\lambda_{\rm s}$, it is still the most important although less 
dominating.  We therefore start by building a {\sffamily Random Forest} 
regressor $y={\rm RF}_1({\rm PC_{MAH,1}},\Mhalo)$, where $y$ is one of the three 
structural quantities. Again, the inclusion of $\Mhalo$ is motivated 
by the fact that halo mass is traditionally considered as one 
of the most important quantities distinguishing halos.  
We also calculate the performance $R_1^2$ as well as the contributions 
$\mathcal{I}_1(x)$ of the regressor. 
We then add the two environmental quantities, the bias factor $b$ 
and the tidal anisotropy $\alpha_\mathcal{T}$, to build a second regressor, 
$y={\rm RF}_2({\rm PC_{MAH,1}},\Mhalo, b, \alpha_\mathcal{T})$, and 
obtain the corresponding $R_2^2$ and $\mathcal{I}_2(x)$. 

\begin{figure*}
	\centering
	\includegraphics[width=12cm]{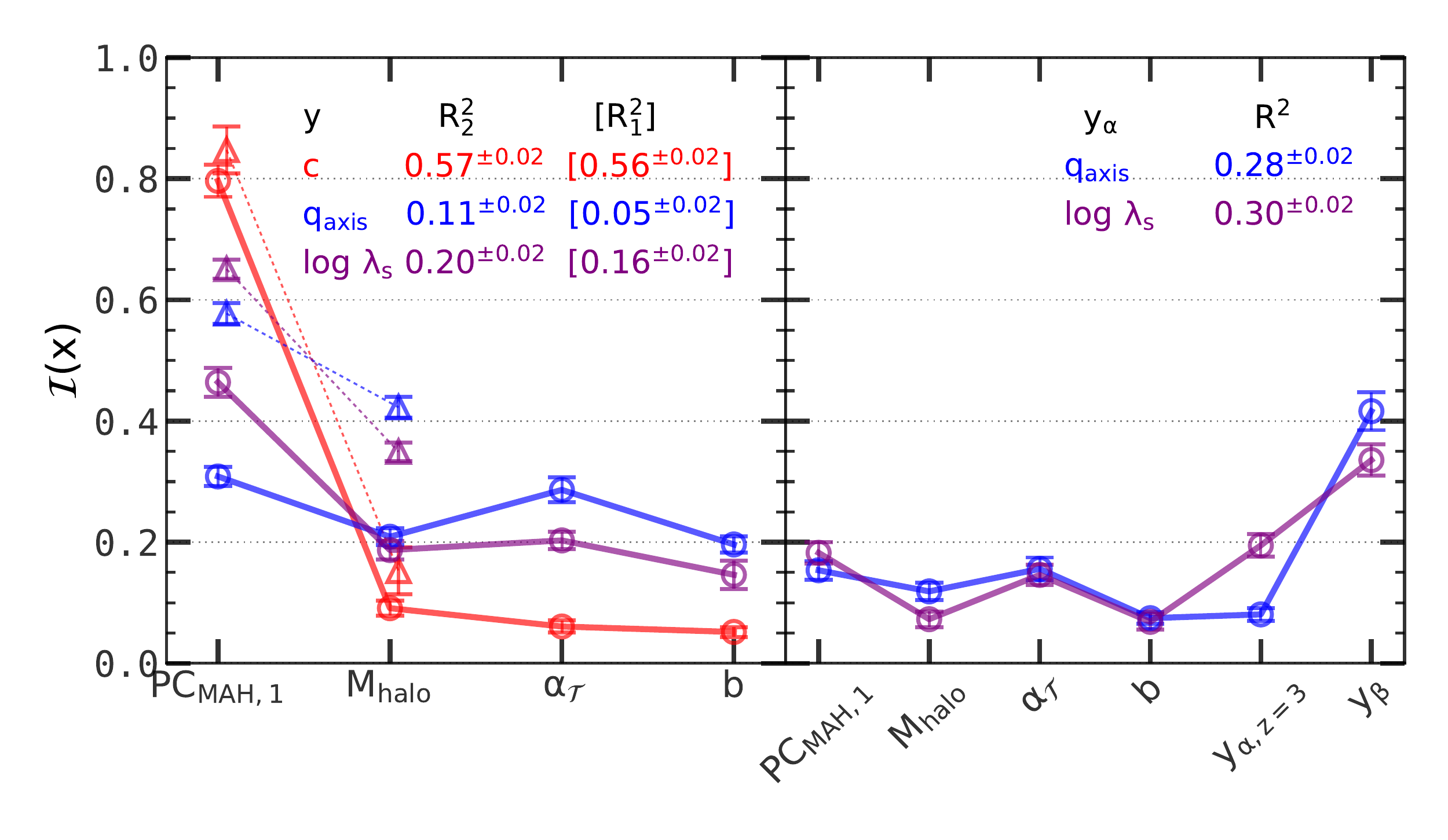}
	\caption{ 
		{\bf Left:} contributions $\mathcal{I}(x)$ from various predictors $x$ 
		to halo structural quantities $y$: concentration $c$ ({\bf red}), shape parameter $q_{\rm axis}$ ({\bf blue}), or the spin parameter $\log \lambda_{\rm s}$ ({\bf purple}). 
		{\bf Solid} lines connecting circles show the results when using the 
		first MAH PC $\rm PC_{MAH,1}$, halo mass $\Mhalo$, tidal anisotropy $\alpha_\mathcal{T}$ and bias factor $b$ as predictors. 
		{\bf Dashed} lines connecting triangles show the results when 
		using only $\rm PC_{MAH,1}$ and $\Mhalo$. 
		The overall performances $R_2^2$ (with environment) and $R_1^2$ 
		(without environment) are also indicated in the panel.
		{\bf Right: }
		similar to the left panel, except that we add the initial condition of each structural property at redshift $z=3$ and a $\beta$-structure parameter $y_\beta$ into the predictors. The target variable, now denoted as $y_\alpha$, is either the shape parameter $q_{\rm axis}$ ({\bf blue}) or the spin parameter $\log \lambda_{\rm s}$ ({\bf purple}). For $q_{\rm axis}$, the $y_\beta$ is $\log \lambda_{\rm s}$, and for $\log \lambda_{\rm s}$, the $y_\beta$ is $q_{\rm axis}$.
		The overall performance $R^2$ is also indicated in the panel.
		In both panels, halos are taken from the mass-limited sample $\rm S_c'$ with $\Mhalo \ge 5\times 10^{11}\msun$ (see \S\ref{ssec_simlation} and Table~\ref{tab_def_samples}). The error bars are calculated using
		100 half-size resamplings without replacement.  }
	\label{fig:struct_vs_other}
\end{figure*}

The left panel of Figure~\ref{fig:struct_vs_other} shows the contribution, $\mathcal{I}(x)$, 
of these two regressors for each of the three structure properties 
described above, with the values of $R^2$ indicated, using 
sample $\rm S_c'$ (see \S\ref{ssec_simlation} and Table~\ref{tab_def_samples}). 
To estimate the uncertainty in the results, we generate 100 random 
subsamples, each consisting of half of the halos randomly  
selected from the original sample $\rm S_c'$ without replacement.
The errors of $R^2$ and $\mathcal{I}(x)$ are then estimated 
as the standard deviations among these subsamples. The reason why we do 
not use the standard bootstrap technique is that the sampling with 
replacement will lead to an artificially large $R^2$ in the {\sffamily Random Forest} 
regressor, because the repeated data points may appear both in the 
training set that shapes the decision trees and the out-of-bag (OOB) 
set (see Appendix~\ref{ssec_method_rf}) that is used to 
test the performance, making the performance overestimated.

In the case of $c$, the inclusion of environment only increases 
$R^2$ from $0.56$ to $0.57$, and the importance value 
$\mathcal{I}(x_{\rm env})$ is below $0.1$. These two results 
indicate that environment has little impact on halo 
concentration $c$, that the concentration $c$ is well 
determined by the first PC of the MAH, and that the weak dependence 
of $c$ on environment is mainly through the dependence of 
halo assembly on environment (assembly bias). 
These are consistent with the finding of
\citet{LuYu:MoHJ:2006:HaloDensityProfileOnFastSlowAccre} 
that the density profiles of individual halos can be modeled 
accurately from their MAHs. 
These are also consistent with the result obtained with the
Gaussian process regression by 
\citet{HanJiaxin:LiYin:2019:GPR_halobias_vs_others}, who found that 
the dependence of halo bias on halo concentration is mainly through 
the dependence of the bias on formation time.

For $q_{\rm axis}$, the $R^2$ 
is doubled, from $0.05$ to $0.11$, when environment factors are 
included. From the importance curve, $\mathcal{I}(x)$, one can 
see that these environment factors take away about half of the 
contribution from the halo mass and $\rm PC_{MAH,1}$. 
These results together suggest that environment can  
affect halo shape significantly and is at least as important as halo mass 
and the PCs of the MAH. 

For the spin parameter, the value of $R^2$ increases from 
$0.16$ to $0.20$ when environment factors are included.
The increase, $0.04$, is about $20\%$, suggesting that these 
environment factors do have a sizable effect on halo spin.
The contribution, $\mathcal{I}(b)+\mathcal{I}(\alpha_\mathcal{T}) = 35\%$, 
is larger than the $20\%$ they contribute to $R^2$, 
suggesting that some of the contribution is actually taken 
from the halo assembly history and halo mass. 
Thus, when interpreting the dependence of halo spin on environment,
one should remember that part of it may actually come from its 
degeneracy with halo assembly history.

Another difference between $c$ and the other two structural parameters 
is in their values of $R^2$. Both $q_{\rm axis}$ and $\log \lambda_{\rm s}$ 
have fairly small $R^2$, much smaller than $50\%$.
This indicates that the major contributors to these two structural 
parameters are not yet found. In general, factors that can affect halo 
structural properties can be classified into three categories: the 
initial conditions of halos, the intrinsic properties of halos 
(e.g., $\Mhalo$, $\rm PC_{MAH,1}$), and halo environment 
(e.g., $\alpha_\mathcal{T}$ and $b$). The interaction of halos with 
their environment depends not only on the environment, but also on 
halo properties. For example, because a halo is coupled to the 
local tidal torque only through its quadrupole, we expect that 
the spin and shape of a halo are correlated. 
Motivated by this and the similar behavior of the 
shape and spin parameters
revealed in~\S\ref{ssec_struct_to_assembly}, we add the spin 
parameter into the predictors of the shape, and vice versa. 
We denote the target structure parameter 
as $y_\alpha$ and the added structure parameter $y_\beta$.   
In addition, we also consider the ``initial condition" of a halo  
by tracing all of the halo particles back to redshift $z=3$
and using these particles to calculate the corresponding shape and 
spin parameters. This quantity for $y_{\alpha}$, denoted 
by $y_{\alpha,z=3}$, is also added into the set of predictors. 
The right panel of Figure~\ref{fig:struct_vs_other} 
shows the result when this set of variables are used to predict halo structures. 
Among all the predictor variables, $y_\beta$ is the most important
for $y_\alpha$, indicating that $q_{\rm axis}$ and $\log \lambda_{\rm s}$
are strongly correlated. For $\lambda_{\rm s}$, its initial value also 
matters, ranked as the second important predictor.
The performance, $R^2$, for both $\lambda_{\rm s}$ and $q_{\rm axis}$
is now boosted significantly, to $\sim 0.3$. 
However, even in this case, $R^2$ is still less than $50\%$, 
meaning that the causes of the main parts of the variances in both  
$\lambda_{\rm s}$ and $q_{\rm axis}$ are still to be identified.  
This also indicates that, unlike the concentration parameter $c$, which 
is determined largely by halo assembly,  $\lambda_{\rm s}$ and $q_{\rm axis}$
may depend on the details of the initial conditions, assembly, and environment.
\cite{YuMorinaga:2020:FilamentaryAccretion} provided a possible scenario where
accretion from filaments may partly account for halo shape and orientation. 
However, the large scatter and weak trend in their results indicate 
that the driving factor of halo shape and orientation is still 
missing. All these suggest that many nuanced factors can contribute
to the variances of $\lambda_{\rm s}$ and $q_{\rm axis}$. 
Models for distributions of $\lambda_{\rm s}$ and $q_{\rm axis}$
have to take into account these nuances by assuming some random 
processes, such as a normal or log-normal process according to the 
central-limit theorem. 

\subsection{Dependence on Halo Mass} 
\label{ssec_mass_on_structure}

\begin{figure*}
	\centering
	\includegraphics[width=14cm]{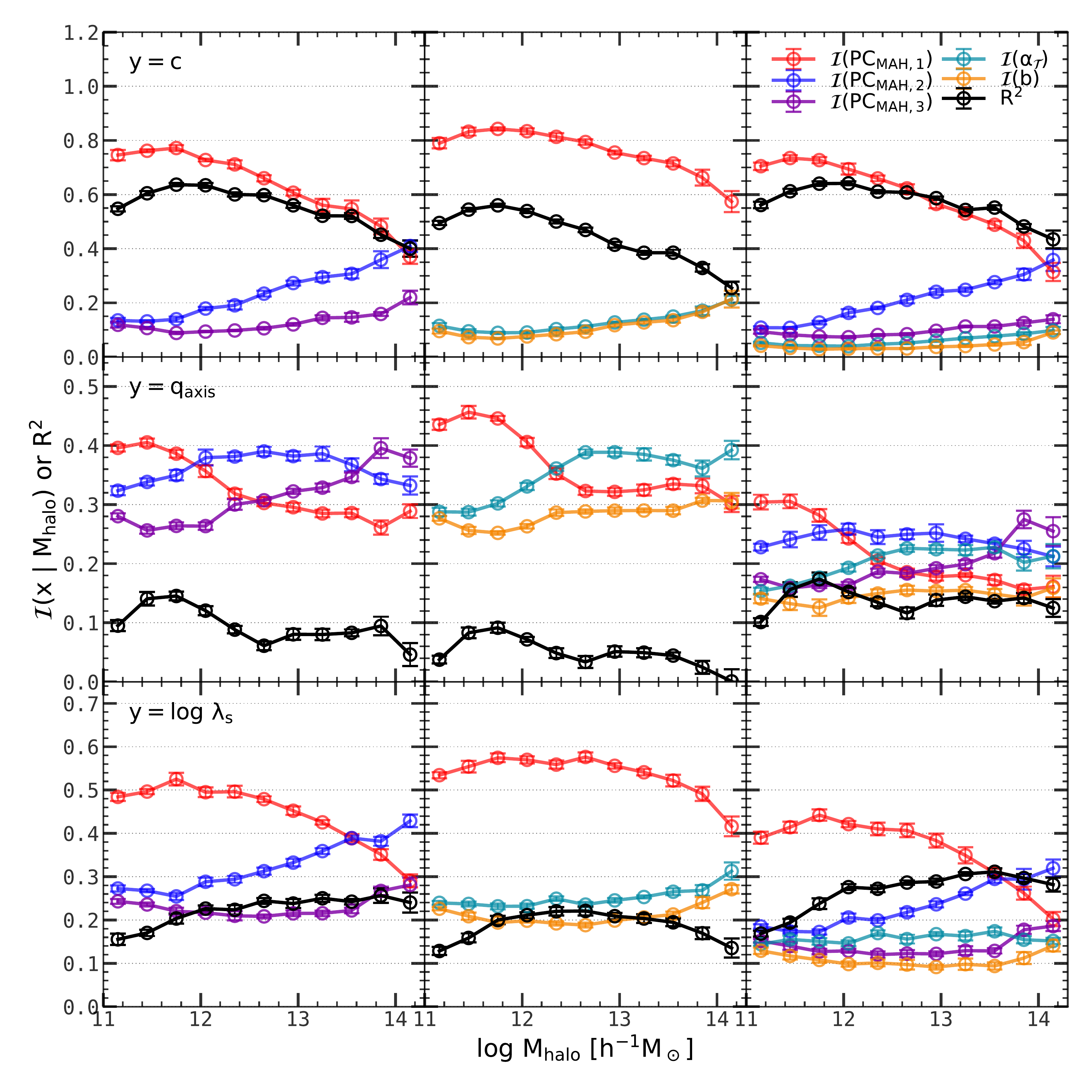}
	\caption{
		Contributions $\mathcal{I}(x)$ from different predictor variables $x$ 
		to the halo structural properties $y$ for halos of different masses. 
		The {\bf upper}, {\bf central}, and {\bf bottom} rows show the results 
		for structural properties $y=c$, $q_{\rm axis}$ and $\log \lambda_{\rm s}$, 
		respectively. The {\bf left}, {\bf middle}, and {\bf right} columns show 
		the regressors built with predictor variables $\bfrm[x]=({\rm PC_{MAH,1}}, 
		{\rm PC_{MAH,2}}, {\rm PC_{MAH,3}})$, 
		$({\rm PC_{MAH,1}}, \alpha_{\cal T}, b)$, and 
		$({\rm PC_{MAH,1}}, {\rm PC_{MAH,2}}, {\rm PC_{MAH,3}}, \alpha_{\cal T}, b)$, 
		respectively.
		Contributions from different predictor variables $x$ are shown with different 
		colors, as indicated in the upper right panel. 
		The overall performance $R^2$ for each regressor is represented by 
		the black symbols. The error bars are obtained by 10 
		half-size resamplings without replacement. Halos are those
		in the large sample $\rm S_L$ (see \S\ref{ssec_simlation} and Table~\ref{tab_def_samples}).
	}
	\label{fig:mass_on_struct}
\end{figure*}	

The regressors for halo structure (presented in 
\S\ref{ssec_struct_to_assembly} and \S\ref{ssec_struct_to_other})
are all built using the mass-limited sample. 
The model training processes and performance measurements 
are thus dominated by low-mass halos, which are more abundant.
However, halos with different masses may have different 
properties. For example, the halo assembly bias, as reflected 
by the correlation between halo formation time and the bias factor, 
is found to be significant only for low-mass halos 
\citep[e.g.,][]{GaoL:SpringelV:2005:AssemblyBias,
GaoL:2007:HaloAssemblyBiasOnSpinOrSub,LiYun_MoHoujun_2008_HaloFormationTimesDef}. 
It is, therefore, interesting to see how the structural properties 
of massive halos depend on assembly and environment, and how 
environmental and assembly effects on these halos are related to 
each other.

Here we quantify such a halo mass dependence by building RF regressors 
for subsamples of a given halo mass, using the large sample $\rm S_L$ 
(see \S\ref{ssec_simlation} and Table~\ref{tab_def_samples}). 
For halos with given mass, $\Mhalo$, and for each of the three structural 
properties, $y=c$, $q_{\rm axis}$ and $\log \lambda_{\rm s}$, we build 
three forest regressors $y={\rm RF}(\bfrm[x]|\Mhalo)$ with different 
sets of predictor variables $\bfrm[x]$: 
\begin{itemize}[leftmargin=*,itemsep=0pt,parsep=0pt, topsep=0pt]
    \item $\bfrm[x]=({\rm PC_{MAH,1}},\, {\rm PC_{MAH,2}},\, {\rm PC_{MAH,3}})$,
    the first three PCs of the halo MAH;
    \item $\bfrm[x]=({\rm PC_{MAH,1}},\, \alpha_{\cal T}, b)$, the 
    first PC of the halo MAH and environmental parameters;
    \item $\bfrm[x]=({\rm PC_{MAH,1}},\, {\rm PC_{MAH,2}},\, {\rm PC_{MAH,3}},\, 
    \alpha_{\cal T},\, b)$, the first three PCs of the MAH plus the environmental 
    parameters.
\end{itemize}
The reason for including ${\rm PC_{MAH,2}}$ and ${\rm PC_{MAH,3}}$ is 
that the MAHs of massive halos may be more complicated than low-mass 
ones, and high-order PCs may be needed to capture the more subtle 
components in their MAHs.

Figure~\ref{fig:mass_on_struct} shows the contribution curves and  
performances of regressors $y={\rm RF}(\bfrm[x]|\Mhalo)$ for different 
halo structural properties, $y$, using different predictor variables, 
$\bfrm[x]$, and for halos of different masses. In the case where 
$\bfrm[x]$ is the first three MAH PCs (panels in the left column), 
the $\rm PC_{MAH,1}$ is always the most important for the structures of 
low-mass halos. However, as the halo mass increases, the importance of 
$\rm PC_{MAH,1}$ decreases and eventually is taken over 
by higher order PCs (the second or the third). 
Massive halos may have more diverse accretion 
histories~\citep[see, e.g.,][who found that tree entropy 
increases with halo mass]{DanailObreschkow_2019_halo_tree_entropy}, 
so their structural properties may also be more complex. 
By using higher order PCs, the complex formation history can be 
captured so that a better prediction for halo structure properties can be 
achieved.

As discussed in \S\ref{ssec_struct_to_other}, for the total halo population, 
environmental effects are different for the three halo structural 
properties. Similar conclusions can be reached for halos of a given mass. 
The middle and right columns in Figure~\ref{fig:mass_on_struct} show 
results for regressors that combine the MAH and environment as 
predictors. For the halo concentration, the PCs of MAH always 
outperform environmental quantities, although there is a 
slight increase in $\mathcal{I}(x)$ for environment quantities 
at the high-mass end. Compared to the regressor with only MAH 
PCs (upper left panel), the performance $R^2$ including 
the two environmental properties (upper right panel) 
only increases slightly, indicating again that the  
environmental effect on halo concentration is mainly through the 
dependence of halo MAH on environment. 

The environmental effect on the shape parameter, $q_{\rm axis}$, 
is totally different. As seen from the contribution curves, the 
environment is as important as MAH, and including the environment 
variables increases $R^2$ significantly. This implies that the
environmental effect on the halo shape parameter is important, 
and that the effect is not degenerate with that of the MAH. 
The environmental contribution to the spin parameter, $\lambda_{\rm s}$, 
is intermediate, larger than that to the concentration 
parameter but smaller than that to the shape parameter. 
The value of $R^2$ after including environment variables increases, 
but less significantly than in the case of the shape parameter. 
This suggests that the environmental effect does contribute to 
halo spins, but part of the contribution is taken from the 
assembly.

\subsection{Halo Assembly Bias}
\label{ssec_env_assembly_bias}

\begin{figure*}
	\centering
	\includegraphics[width=12cm]{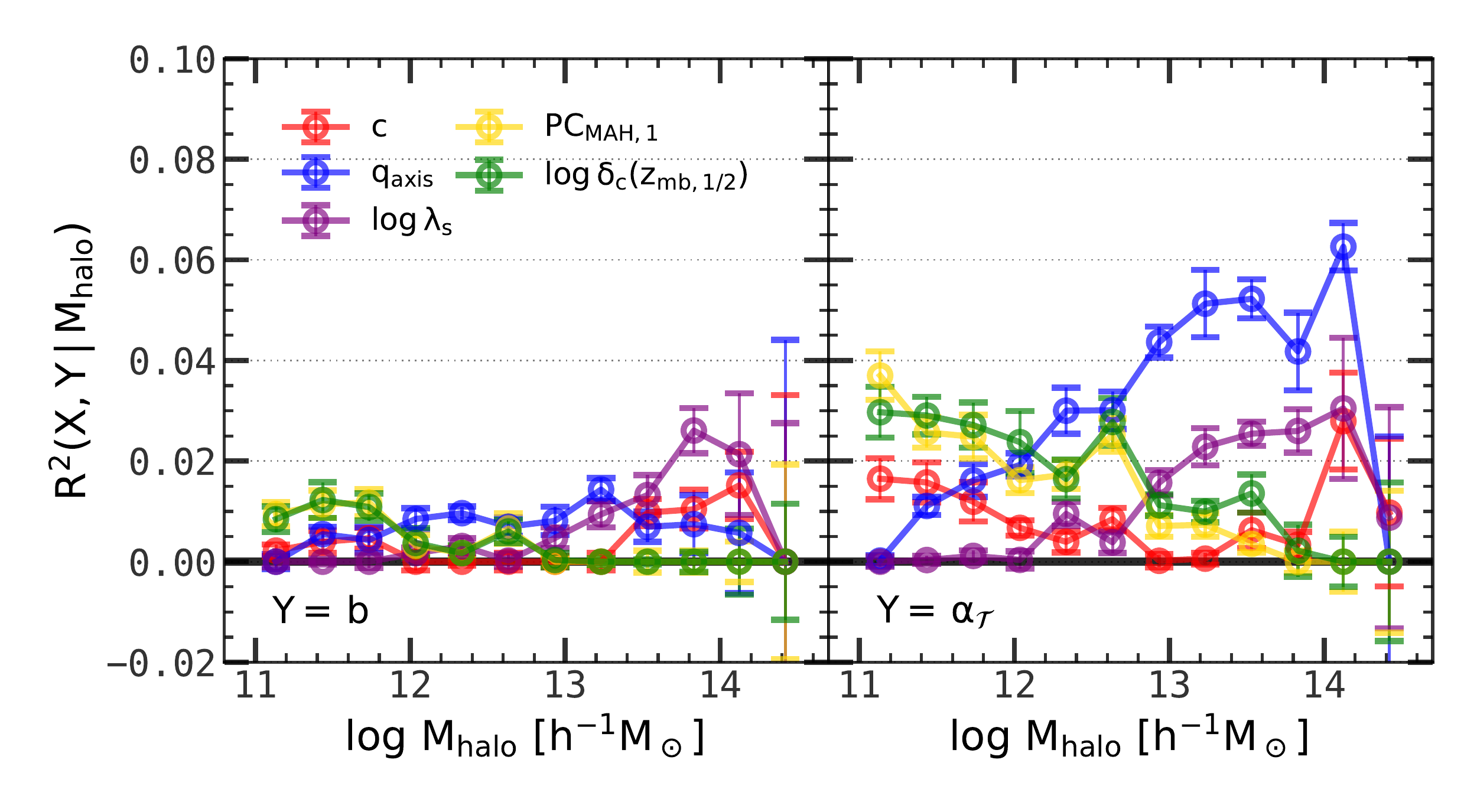}
	\caption{
		{\sffamily Random Forest} regression performance $R^2(X,Y|\Mhalo)$ of pairs of variables $(X,\,Y)$ for halos with a given mass, where $Y$ is an environmental quantity ({\bf left} panel: halo bias factor $b$; {\bf right} panel: tidal anisotropy parameter $\alpha_\mathcal{T}$), and $X$ is either a structural quantity or an assembly history quantity, as represented with different colors. The error bars of $R^2$ are obtained by 10 half-size resamplings without replacement. Halos are taken from the large sample $\rm S_L$ (see \S\ref{ssec_simlation} and Table~\ref{tab_def_samples}).
	}
	\label{fig:assembly_bias}
\end{figure*}	

As a final demonstration of the application of 
the {\sffamily Random Forest} regressor, 
we show how assembly parameters correlate with environment for halos of a given 
mass. Such a correlation is usually referred to as the 
halo assembly bias. The purpose here 
is to identify the best correlated pair of variables $(X,\,Y)$ at given halo mass, 
where $X$ is an assembly property and $Y$ is an environmental property. The method 
to measure the correlation strength is straightforward. First, we bin halos in 
sample $\rm S_L$ (see \S\ref{ssec_simlation} and Table~\ref{tab_def_samples}) 
into subsamples according to the halo mass. Within each 
subsample, we build a {\sffamily Random Forest} regressor for each pair of variables 
$(X,\,Y)$. The value of $R^2(X,Y|\Mhalo)$ then provides a measurement of 
the correlation strength between the two quantities.  
Figure~\ref{fig:assembly_bias} shows the results for different cases, 
where the environmental quantity $Y$ is either the bias factor $b$ or the 
tidal anisotropy parameter $\alpha_\mathcal{T}$, and $X$ is either 
$\rm PC_{MAH,1}$ or $\log \delta_{\rm c}(z_{{\rm mb},1/2})$.
As one can see, the dependence of $\log \delta_{\rm c}(z_{{\rm mb},1/2})$ 
on $b$ is present only for halos with $M_{\rm halo}<10^{13}h^{-1}{\rm M}_\odot$,  
and totally absent for more massive ones.
The values of $R^2$ between $b$ and the two assembly properties 
are both smaller than $2\%$, indicating that the correlation between $b$ and 
assembly history is weak. These results are consistent with those 
obtained previously \citep[e.g.,][]{GaoL:SpringelV:2005:AssemblyBias, 
GaoL:2007:HaloAssemblyBiasOnSpinOrSub,LiYun_MoHoujun_2008_HaloFormationTimesDef}:
the assembly bias is significant only for low-mass halos, and  
one has to average over a large number of halos to 
detect the weak trend. 

For comparison, we also build regressors between structure properties
($c$, $\lambda_{\rm s}$ and $q_{\rm axis}$) and $b$ for halos
of a given mass. The results are shown in the left panel of Figure~\ref{fig:assembly_bias}. 
Clearly, the dependence of these properties on $b$ is also weak. 
The results are consistent with those obtained previously by 
\citet{MaoYaoyuan:2018:SecondaryBias}, who found 
that $b$ is better correlated with $c$ and $\lambda_{\rm s}$ than 
with assembly properties for massive halos.

In a recent paper, \cite{RamakrishnanS:2019:CosmicWebAnisotropy:PrimaryBias} 
showed that the tidal anisotropy parameter, $\alpha_\mathcal{T}$, is a good 
variable that correlates well with many structural properties. 
We present the correlation between $\alpha_\mathcal{T}$ 
and other halo quantities in the right panel of 
Figure~\ref{fig:assembly_bias}. It is clear that $\alpha_\mathcal{T}$ 
shows a better correlation with halo intrinsic properties than the bias factor, 
as indicated by the larger values of $R^2$. In particular, 
the correlations between the assembly properties 
($\delta_{\rm c}(z_{\rm mb, 1/2})$ and $PC_{\rm MAH,1}$)
and $\alpha_\mathcal{T}$ are significant, except at the very massive end. 
In \S\ref{ssec_struct_to_other}, we demonstrate that part of the contribution from 
the environment to the structural properties is produced by the degeneracy 
between the environment and the halo assembly history. The strong 
relation between $\alpha_\mathcal{T}$ and 
assembly history is a proof of this degeneracy. Note that $q_{\rm axis}$ 
shows a strong correlation with $\alpha_\mathcal{T}$ for massive halos, 
indicating that the local tidal field plays an important role in determining 
the shape of the halo. 

\begin{figure*}
	\centering
	\includegraphics[width=\linewidth]{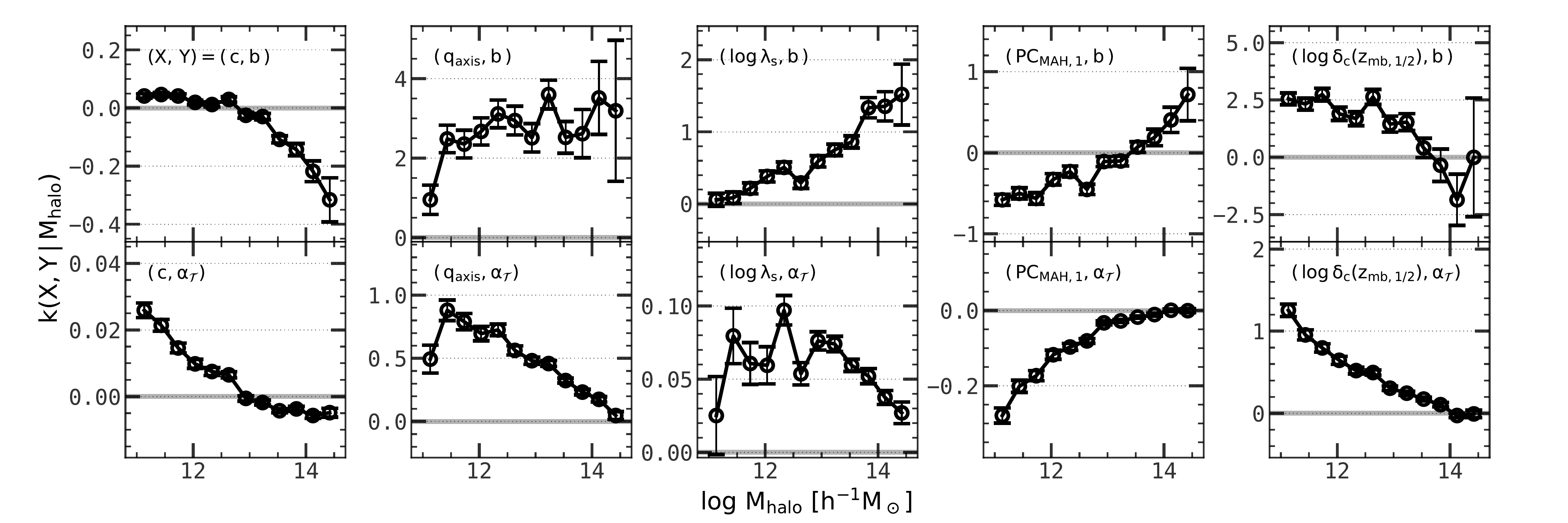}
	\caption{
		Linear regression slope $k(X,Y|\Mhalo)$ of pairs of variables $(X,\, Y)$ for halos with a given mass, where $Y$ is an environmental quantity and $X$ is either a structural quantity or an assembly history quantity. Each panel is for a pair of $(X,\,Y)$. The error bars are standard errors estimated from linear regression model. Halos are taken from the large sample $\rm S_L$ (see \S\ref{ssec_simlation} and Table~\ref{tab_def_samples}).
	}
	\label{fig:assembly_bias_linear_regress}
\end{figure*}	

As mentioned above, the small values of $R^2$ between assembly and environment
imply that assembly bias is a weak relation compared to the variance.   
Detecting such a bias thus needs the average over a large number of 
halos. To obtain the mean trend in the bias relation, we can build linear 
regression models between pairs of variables and use the slopes of the 
regression, $k$, to represent the mean trend between the two variables in 
question. 
Figure~\ref{fig:assembly_bias_linear_regress}
shows the slopes of halo properties with $b$ and $\alpha_{\cal T}$ for 
halos with different masses. 
\myrevise[Given any two variables, a larger absolute value of 
$k$ means a more significant linear correlation of 
the two variables for halos of a given mass.
The relations between halo environment and different halo properties 
show different trends with halo mass. 
For example, the slope $k(\log\delta_{\rm c}(z_{{\rm mb},1/2}), b|\Mhalo)$ 
is significant only at the low-mass end, 
while $k({\rm PC_{MAH,1}}, b|\Mhalo)$ is significant at 
the low-mass end and becomes less significant at the high-mass 
end, suggesting that halo assembly bias is more important for 
halos of lower mass. 
In contrast, the slopes $k(c, b|\Mhalo)$, $k(q_{\rm axis}, b|\Mhalo)$ 
and $k(\log \lambda_{\rm s}, b|\Mhalo)$ are all more 
significant at the high-mass end.]  As discussed 
in \citet{WangHuiyuan_MoHoujun_2011_HaloDependOnEnv}, 
this halo mass dependence is a result of the competition
between the environmental effect and the self-gravity of the 
halo. For example, a higher density environment not only provides 
more material for halos to accrete, but also is the location of a 
stronger tidal field that tends to suppress halo accretion. 
Depending on the halo mass, one of the two effects dominates.
For example, for lower mass halos where self-gravity is weaker, 
the local tidal field may be more effective in making them more elongated
and spin faster, and in preventing them from accreting new mass, 
so as to make their mass assembly earlier and concentration 
higher. This is indeed what can be seen from the lower panels of 
Figure~\ref{fig:assembly_bias_linear_regress}.  
We note, however, that the physical units of the different curves in 
these panels are not the same, so that these curves cannot 
be compared directly with one another. Note also that a 
significant value of $k$ does not necessarily imply  
a significant $R^2$ in Figure~\ref{fig:assembly_bias}, 
and vice versa, because a linear model may not
represent faithfully the data of nonlinear relations. 
In addition, the value of $R^2$ depends not only 
on the value of $k$, but also on the absolute value of 
the variance in the regressed sample.

\section{Summary and discussion}
\label{sec_summary}

In this paper, we have used the ELUCID $N$-body simulation to 
relate the structural properties of dark matter halos to their assembly history 
and environment. Our analysis is based on the PCA and the RF regressor. 
Our main results and their implications can be summarized as follows.

First, PCA is a simple and yet effective tool for reducing the dimension of 
the halo MAH, and it is preferred over formation times in characterizing the 
halo MAH. It has the following three major advantages 
(see \S\ref{sec_assembly} and \S\ref{ssec_struct_to_assembly}):
\begin{itemize}[leftmargin=*,itemsep=0pt,parsep=0pt, topsep=0pt]
	\item PCs are complete and linearly independent. The first three PCs 
	can already explain more than $80\%$ of the variance of the halo MAH, 
	with a reconstruction error $<10\%$.
	\item The PCs of the MAH have clear physical meanings. The lower order 
	PCs, such as the first PC, are tightly related to the (halo formation) 
	times when a halo forms fixed fractions of its current mass, 
	such as $z_{{\rm mb}, 1/2}$, $z_{{\rm mb}, 0.04}$, and so on. 
	Higher order PCs, on the other hand, are related to more subtle events, 
	such as the presence of major mergers.
	\item PCs are the best, among all assembly indicators, to explain 
	halo structure. The first PC, among all assembly indicators, accounts 
	for about $67\%$, $20\%$, and $28\%$ of the variance in 
	halo concentration, shape parameter, and spin parameter, respectively.
\end{itemize}

Second, the dependence on assembly and environment is quite  
different for the three halo structural properties 
(see \S\ref{ssec_struct_to_other}). About $60\%$, $10\%$, and $20\%$ 
of the variances in $c$, $q_{\rm axis}$, and $\lambda_{\rm s}$, 
respectively, can be explained by four predictors: 
$\rm PC_{MAH,1}$, $\Mhalo$, $\alpha_{\cal T}$ and $b$.
Halo concentration is dominated by the first PC of the MAH, 
with the contributions from other factors negligible.
For $q_{\rm axis}$ and $\lambda_{\rm s}$, there is no single 
property of assembly that is dominating. The environment has 
significant effects on these two structural parameters, but 
its effect on $\lambda_{\rm s}$ is degenerate with the assembly history.
The correlation between $q_{\rm axis}$ and $\lambda_{\rm s}$ 
is strong, indicating that these two quantities share some 
common origins. The initial condition is also important for 
$\lambda_{\rm s}$. However, putting all the factors together, 
we see that the values of $R^2$ are still smaller than $0.5$ for both 
$q_{\rm axis}$ and $\lambda_{\rm s}$, indicating that 
these two halo quantities may be affected by many subtle 
factors and thus are difficult to model. 

Third, the structural properties depend mainly on
the first PC of the MAH for low-mass halos, but have a significant 
dependence on higher order PCs for high-mass halos. The conclusions 
for the overall population still hold for halos of a 
given mass: environment has almost no effect on $c$ once the MAH is 
included; environment is more important for $q_{\rm axis}$
and $\lambda_{\rm s}$, although its effect on $\lambda_{\rm s}$ is 
partly degenerate with that of the MAH.

Fourth, the tidal anisotropy, $\alpha_{\cal T}$, has a stronger correlation
with halo assembly and structure than the bias factor $b$ does. 
We see that $\alpha_{\cal T}$ is correlated with halo assembly history for 
all halos except the most massive ones, and it also shows a 
significant correlation with $q_{\rm axis}$, indicating 
that the local tidal field plays an important role in shaping 
a halo. However, all types of assembly bias tested here are weak
compared to the variance in the relation, and averaging with a large 
sample is needed to detect them reliably .

For dimension-reduction tasks such as those for the halo MAH, one may also 
use nonlinear algorithms, such as the locally linear embedding 
\citep[LLE;][]{RoweisS2000_LLE}, and the spectral embedding 
\citep{BelkinM:2003:SpecEmbedding}. The degrees of freedom of these 
manifold-learning techniques are higher than those of the linear 
algorithms, so that they are not stable for noise data. 
Indeed, we have tried implementing LLE in halo MAH, but we found that it 
does not outperform the PCA in terms of both the reconstruction 
error and the correlation with halo structural properties.

The correlation of structural properties with halo assembly and 
environment revealed by the RF analysis provides insights 
into the halo population in the cosmic density field.  
Since galaxies form and evolve in dark matter halos, understanding 
the formation, structure, and environment of dark matter halos 
is a crucial step in establishing the link between galaxies and halos. 
Empirical approaches, such as  
(sub)halo abundance matching \citep{MoHJ:MaoS:WhiteSDM:1999:LBG:AM, 
Vale2004, GuoQi:2010:SHAM}, 
clustering matching \citep{GuoHong_ZhengZheng_2016_SHAM_SCAM}, 
age matching \citep{HearinAP:2013:SubhaloAgeMatching}, 
conditional color-magnitude diagrams \citep{XuHJ:2018:CCMD}, 
halo occupation distributions \citep{JingYP:1998:HOD, BerlindAA:2002:HOD}, 
conditional luminosity functions \citep{YangXiaohu:2003:CLF_M2LRatioTurnOver}, 
and those based on star formation rate \citep{LuZhankui:2014:EmpiricalModel,
MosterBenjaminP_2018_EmpiricalModel,BehrooziPeter:2019:UniverseMachine}, 
all use halo properties to make predictions for the galaxy population.   
A key question in all of these models is which halo 
quantities should be used as the predictors of galaxies. 
Using too little information about the halo population
will make the model too simple to capture the real effects of 
halos on galaxy formation; using too many halo properties 
may be unnecessary because of the degeneracy between them. 
Our results, therefore, provide a foundation for building 
galaxy formation models such as those listed above.

For readers who are interested in generating Monte Carlo samples
of halos of different structural properties, including  
dependencies on assembly and environment properties, 
we provide both an online calculator and a programming interface at 
\url{https://www.chenyangyao.com/publication/20/haloprops/}.

\section*{Acknowledgements}
This work is supported by the National Key R\&D Program of China
(grant Nos. 2018YFA0404502, 2018YFA0404503), and the National 
Science Foundation of China (grant Nos. 11821303, 11973030, 
11761131004, 11761141012, 
\myrevise[11833005, 11621303, 11733004, 11890693, 11421303] ). 
Y.C. gratefully acknowledges 
the financial support from China Scholarship Council.

\appendix

\section{Methods of analysis} 
\label{sec_method}

Throughout this work, we use two statistical methods to analyze halo properties. 
The PCA is used to reduce the complexity
of quantities in high-dimensional space, and the EDT, also called the RF, is used to study 
correlations among different quantities. A brief description of the two 
methods is given below.
For a more detailed description, see Pattern Recognition and Machine Learning by~\citet{BishopC:2006:PRML}. 
The programming interfaces and implementation can be found in 
\href{https://scikit-learn.org/stable/modules/ensemble.html}{scikit-learn}.

\subsection{Principal Component Analysis} 
\label{ssec_method_pca}

PCA is an unsupervised, reduced linear Gaussian dimension-reduction method 
\citep{PearsonKarl:1901:PCA, HotellingH:1933:PCA}.
Consider a set of $\rm N$ vectors $\bfrm[X] = 
(\bfrm[x] _1,\ ...,\ \bfrm[x] _{\rm N})^{\rm T}$, 
each in an $\rm M$-d space, $V_{\rm M}$. The idea of the PCA is to find an 
$\rm M'$-d subspace, $V_{\rm M'}$ ($\rm M' \le M$), in which the projection of 
$\bfrm[X]$, 
\beq 
\bfrm[X']=\bfrm[X]\bfrm[P]
\eeq 
has maximal variance, where $\bfrm[P]$ is the projection operator.
It can be shown that the problem to be solved is equivalent to solving the 
eigenvalue problem for the sample covariance matrix of $\bfrm[X]$, defined as 
\beq
\bfrm[S] = \sum_{i=1}^{\rm N}
(\bfrm[x]_i-\overline{\bfrm[x]})(\bfrm[x]_i-\overline{\bfrm[x]})^{\rm T},
\eeq
where $\overline{\bfrm[x]}=\sum_{i=1}^{\rm N}\bfrm[x]_i/{\rm N}$ is the sample mean. 
If we rank the eigenvalues in a descending order, the first eigenvector of 
$\bfrm[S]$, $\bfrm[v]_1$, is the direction along which the sample 
has the maximum projected variance, $\lambda_1=\bfrm[v]_1^{\rm T} \bfrm[S] \bfrm[v]_1$.
This variance is exactly the first eigenvalue of $\bfrm[S]$. 
Similarly, the $i$th eigenvector and the $i$th eigenvalue are, 
respectively, the direction and value of the $i$th largest variance.
Consider the space $V_{\rm M'}$ spanned by the first $\rm M'$ eigenvectors.
The linearity of the transformation can be used to prove that 
the projected variance in ${\rm M}'$ 
is $\sigma^2=\sum_{i=1}^{\rm M'}\lambda_i$. 
Thus, one can project each data point $\bfrm[x]$ into $V_{\rm M'}$ 
by $\bfrm[P]=(\bfrm[v]_1,\ ...,\ \bfrm[v]_{\rm M'})$: 
$\bfrm[x']=\bfrm[P]^{\rm T}\bfrm[x]$, to find a lower-dimension 
representation for it. The $i$th component of $\bfrm[x']$
is called the $i$th PC of this data point in 
the sample.

In general, any dimension-reduction algorithm will lose information 
contained in the original data. In PCA, the proportional variance 
explained (PVE) by the $i$th PC, defined as
\beq
{\rm PVE}_i = \frac{\lambda_i} { {\rm Tr}(\bfrm[S]) },
\eeq
is used to quantify the importance of the $i$th PC. 
The cumulative PVE, defined as 
\beq
{\rm CPVE_{M'}}=\sum_{i=1}^{\rm M'}{\rm PVE}_i,
\eeq
can be used to quantify the performance of using the first $\rm M'$ PCs
in the dimension reduction. Typically, if the data in question are generated 
from an intrinsic process of lower dimension, the CPVE should quickly converge to 
$1$ as $\rm M'$ increases. We will see that 
halo assembly histories have this property (\S\ref{ssec_assembly_PC}).

The inverse operation of the projection, 
$\tilde{\bfrm[x]}=\bfrm[P]\bfrm[x']$, allows one to 
reconstruct the original vector, but with information loss. 
We use the following quantity, 
\begin{equation}
e=\norm{\tilde{\bfrm[x]}-\bfrm[x]}/\norm{\bfrm[x]},
\end{equation}
to quantify the reconstruction error of the data point $\bfrm[x]$.

\subsection{The Random Forest} 
\label{ssec_method_rf}

The RF regressor or classifier is a supervised, 
decision-tree-based,
highly nonlinear, nonparametric model ensemble method in statistical 
learning \citep{BreimanLeo:2001:RandomForest}. 
Here we first introduce the decision tree algorithm, and then we discuss how 
the trees are combined into a forest to make a regression or classification.

Given a set of observations $D=\{(\bfrm[x]_i, y_i)\}_{i=1}^{\rm N}$, each 
consisting of a vector of predictor variables
$\bfrm[x]_i \in \mathbb{R}^{\rm M}$, and a target random variable 
$y_i$ (continuous in the regression problem, discrete in the classification 
problem), a decision tree $\rm T$ can be trained to fit the data by
minimizing some error functional $\mathcal{E}({\rm T}|D)$. In regression problems,
a common choice for the error functional, which we adopt here, 
is the mean residual sum-of-square,
\begin{equation}
	\mathcal{E({\rm T}|D)} = {1 \over {\rm N}} \sum_{n=1}^{\rm N} 
	[f(\bfrm[x]_i)-y_i]^2,
\end{equation}
where $f(\bfrm[x]_i)$ is the predicted value for the $i$th observation 
by the tree $\rm T$. The tree can then be used to predict the target 
value for a future test observation, $\bfrm[x]$. 

A decision tree is built by sequentially bipartitioning the feature 
space along some axes. At each partitioned region $\mathcal{R}$ in 
the feature space, the tree fits the observations by a constant function, 
\begin{equation}
	f(\bfrm[x])= { 1 \over {\rm N}_\mathcal{R} } \sum_{n=1}^{{\rm N}_\mathcal{R}} y_i,
\end{equation}
where the summation is over all observations 
in the region $\mathcal{R}$, and ${\rm N}_\mathcal{R}$ is 
the number of training observations in this region.
In the first step, the variable $x$ to be 
bipartitioned and the position of the partition plane are both chosen 
to minimize the error functional $\mathcal{E}({\rm T}|D)$. 
After a partition, the feature space is split into two subregions, 
each of which can be bipartitioned further to minimize the error functional. 
This recursive process partitions the feature space into a tree-like 
structure and is continued until some stop criterion 
(e.g. the maximum tree height, or the maximum number of tree nodes, 
or the minimum number of data points in individual leaf nodes, 
defined as the nodes at the top of the tree) is achieved. 

Once a tree is built, the amount of error reduced by partitioning 
variable $x$ can be computed as
\begin{equation}
	\mathcal{I}_{{\rm T}|D}(x) = C \sum_{n} {\rm N}_n \left(\mathcal{E}_n-{ {\rm N}_{n_{\rm L}} \over {\rm N}_n }\mathcal{E}_{n_{\rm L}}-{ {\rm N}_{n_{\rm R}} \over {\rm N}_n }\mathcal{E}_{n_{\rm R}}\right).
\end{equation}
Here the summation is over all tree nodes partitioned by variable $x$. 
$\mathcal{E}_n$ 
is the error functional computed at observations in the region represented by 
node $n$ before partition; $\mathcal{E}_{n_{\rm L}}$ and 
$\mathcal{E}_{n_{\rm R}}$ are the errors computed at the left 
and right child nodes of node $n$, respectively, 
after partition; and $N_n$, $N_{n_{\rm L}}$ and $N_{n_{\rm R}}$ 
are the number of observations in node $n$ and 
its two child nodes, respectively. The normalization factor 
$C$ is chosen so that the summation of $\mathcal{I}$ from all feature 
variables is one. So defined, the quantity $\mathcal{I}_{{\rm T}|D}(x)$ 
is the amount of contribution from variable $x$ in building the 
regressor and can therefore be viewed as the important value 
of the variable in explaining target $y$.

Such a tree model suffers from the overfitting problem:
the more complicated the tree is, the less training error it will 
have. However, the tree will eventually be dominated by noise as 
its height increases. Many methods have been proposed to control such 
overfitting, for example, the cross-validation, bootstrap, and jackknife
ensembles. Here we adopt the RF, an extension of the bootstrap ensemble, 
designed specifically to deal with overfitting in tree-like algorithms. 
The building of an RF involves two levels of randomness. First, 
one uses $n_{\rm re}$ bootstrap resamplings, and each is used to train 
a tree ${\rm T}_i\ (i=1,...,n_{\rm re})$. Second, when training 
each of the $n_{\rm re}$ trees, only a random subset (size $n_{\rm var} < {\rm M}$) 
of all $\rm M$ predictor variables is used at each partition step. 
The $n_{\rm re}$ trees are then combined, and the final prediction for a 
given feature $\bfrm[x]$, denoted as ${\rm RF}(\bfrm[x])$, 
is then averaged among all trees 
(arithmetic average in regression, and majority voting in 
classification). Also, the importance of the predictor, 
$\mathcal{I}_{{\rm T}|D}(x)$, is averaged among trees, which we denote 
as $\mathcal{I}_{{\rm RF}|D}(x)$. The overall performance of the RF 
is represented by 
the explained variance fraction, $R^2$, defined as
\begin{equation}
	R^2 = 1- \frac{ \sum_{n=1}^{\rm N_T} [y_n-{\rm RF}(\bfrm[x_n])]^2 }{ 
	\sum_{n=1}^{\rm N_T} [y_n-\overline{y}]^2 },
\end{equation}
where in principle the summation should be computed over an 
independent test sample, $\rm T$, of size $\rm N_T$. The RF regressor 
has the advantage that the test performance can be directly 
estimated with the OOB sample in the bootstrap process, 
and therefore an extra test sample is not necessary. 
In addition, RF does not suffer from the issue of scaling or arbitrary 
transformation of predictors, which exists in many nonlinear approaches,
such as K-nearest-neighbors (KNNs) and support vector machines (SVMs).

The RF method has some free parameters to be specified, and we choose them 
based on the following considerations: (1) The number of trees in the forest, 
$n_{\rm re}$, should be as large as possible to suppress overfitting. 
But a larger $n_{\rm re}$ is computationally more difficult. 
In our analysis, we choose $n_{\rm re}=100$, which is sufficiently 
large for most applications of RF. (2) The number of predictors randomly 
chosen in the partition, $n_{\rm var}$, also controls the suppression of 
overfitting. We optimize the value of $n_{\rm var}$ by 
maximizing the OOB score through grid searching. 
(3) The tree termination criterion affects 
the complexity of each tree. We choose to control the number of data points 
in the leaf nodes, $s_{\rm leaf}$. This choice makes the tree  
self-adaptive when more data points are available, and also reduces 
issues associated with transformations of target variables and the 
choice of the error functional. The value of $s_{\rm leaf}$ is also 
optimized by maximizing the OOB score, again through grid searching.

\section{Definitions of Halo Formation Times} 
\label{sec_ftime_def}

Because of the diversity in MAHs, different formation times can be 
defined to describe different aspects of the assembly. Here we summarize 
the halo formation times we used in our analysis. Most of the 
definitions can be found in 
\citet{LiYun_MoHoujun_2008_HaloFormationTimesDef}, but we also add some 
new definitions that have been used by others. 
Since the halo mass and redshift in different time steps are 
discrete, the mass-redshift relation is linearly interpolated 
within adjacent time steps.
\begin{itemize}[leftmargin=*, itemsep=0pt, parsep=0pt, topsep=0pt]
	\item $z_{{\rm mb},1/2}$: 
	the highest redshift at which the main branch has 
	assembled half of its final mass $\Mhalo$ 
	\citep[e.g., ][]{vandenBosch:2002:HaloUniversalMAH,
	ShiJJ2018_HaloBimodalFormation}. 
	\item $z_{\rm mb,core}$: 
	the highest redshift at which the main branch has reached 
	a fixed mass $M_{\rm h,\ core}=10^{11.5}\msun$. A halo at such a mass typically 
	has the highest star formation 
	efficiency~\citep{vandenBosch:2003:CLF, YangXiaohu:2003:CLF_M2LRatioTurnOver}, and 
	therefore is capable of forming a bright galaxy in it.
	\item $z_{{\rm mb},0.04}$: 
	this is the highest redshift at which the main 
	branch of a halo assembled $4\%$ of its final mass. This formation time is 
	found to be related to the concentration of the halo in a wide range of 
	cosmological models \citep[see ][]{ZhaoDonghai_2009_HaloConcenAnalytical}.
	\item $z_{{\rm mp},1/2}$: 
	similar to $z_{{\rm mb},\ 1/2}$, but using the most massive     
	progenitors (MMPs) in the entire tree rooted from a halo, instead of the 
	main branch. This definition is used in 
	\citet{WangHuiyuan_MoHoujun_2011_HaloDependOnEnv}.
	\item $z_{\rm mp,core}$: similar to $z_{\rm mb,\ core}$, but using MMPs.
	\item $z_{{\rm 0.02},1/2}$: the highest redshift at which half of the 
	halo mass has been assembled into its progenitors with masses 
	$\ge 2\%$ of the halo mass. This definition is used in  
	\citet{NavarroJ_FrenkC_WhiteS_1997_NFWProfile} to study the correlation between 
	halo concentration and formation history (see also 
	\citet{JeasonDanielAkila:2011:PCA:HaloProperties}, where a different mass 
	threshold is used).
	\item $z_{{\rm core},1/2}$: 
	the highest redshift at which half of the halo mass has 
	been assembled into its progenitors with masses 
	$\Mhalo \geq M_{\rm h,\ core} = 10^{11.5}\msun$. This represents the time when
	the massive progenitors are capable of forming large amounts of stars.
	\item $z_{{\rm core},f}$: the highest redshift at which a fraction 
	$f=\frac{1}{2}(\Mhalo / M_{\rm h,\ core})^{-\gamma}$ of the halo mass 
	has been assembled into its progenitors with masses $\Mhalo \geq M_{\rm h,\ core}$, 
	where $\gamma=0.32$. This definition takes into account 
	the dependence of star formation efficiency on halo mass 
	\citep{YangXiaohu:2003:CLF_M2LRatioTurnOver}.
	\item $z_{\rm mb,vmax}$: the redshift at which the main branch has achieved 
	its maximum virial velocity. This definition, therefore, 
	reflects the formation of the gravitational potential well.
	\item $z_{\rm mp,vmax}$: the same as $z_{\rm mb,\ vmax}$, but using 
	the maximum virial velocity of MMPs.
	\item $z_{\rm lmm}$: the last major merger time of a halo. Here the major 
	merger is defined as a merger event in which the mass ratio 
	$r=m/M$ between the two merger parts is larger than one-third. 
	A major merger is a violent event and may change the halo structure significantly.
	\item $s_{\rm tree,\beta}$: the tree entropy with entropy update efficiency 
	$\beta$ \citep[see][]{DanailObreschkow_2019_halo_tree_entropy}. 
	Different from the formation parameters defined above, this parameter 
	is not associated with any specific event of the halo assembly history, 
	but it describes the complexity of the whole tree. By construction, 
	$s_{\rm tree,\beta}$ is bound to the range $[0,1]$. 
	A close-to-zero $s_{\rm tree,\beta}$ represents a continuous accretion 
	history, while a close-to-one $s_{\rm tree,\beta}$ describes a history that 
	is given by the merger of two progenitors of equal mass. The parameter 
	$\beta \in [0,\,1]$ controls the balance between the entropy inherited 
	from the progenitors and that generated in recent merger events. 
	We choose $\beta = 0.1$ so that the tree entropy can reflect its
	assembly history at high $z$.
\end{itemize}

\begin{figure*}[h!]
	\centering
	\includegraphics[width=12cm]{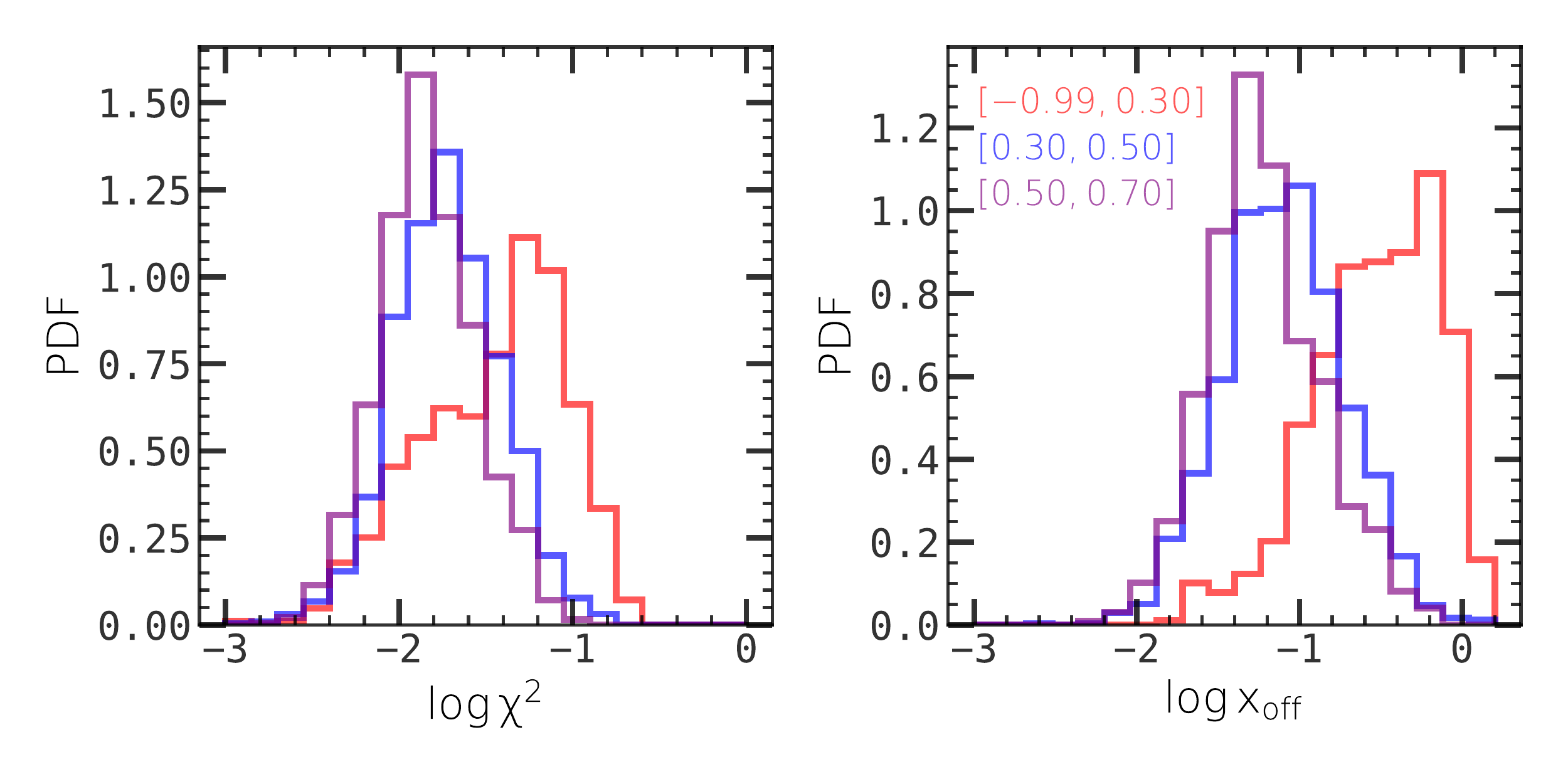}
	\caption{ Distributions of halo relaxation-related parameters. {\bf Left:} relaxation parameter $\chi^2_{\rm c}$, which is the normalized $\chi$-squared value in fitting the NFW profile. {\bf Right:} offset parameter $x_{\rm off}$, which is the normalized distance from the center of mass to the most bound particle of the halo. 
		In each panel, halos are taken from the mass-limited sample $\rm S_{c}'$ with $\Mhalo \ge 5\times10^{11}\msun$ (see \S\ref{ssec_simlation} and Table~\ref{tab_def_samples}), and are divided into three sub-samples according to their last major merger time $\log\,\delta_{\rm c}( z_{\rm lmm} )$, indicated in the right panel with different colors. }
	\label{fig:struct_relaxChi2_xoff_dist}
\end{figure*}

\section{Effect of Unrelaxed Halos}
\label{sec_sim_unrelax}

As demonstrated by~\cite{MacCio2007}, halos that have undergone 
recent major mergers may be unrelaxed and have structural properties
significantly different from virialized halos \citep[see also][]{LudlowAD:2012:HaloConcentrationAndHistoryUnrelaxed}. 
Following~\cite{MacCio2007} we define two parameters to quantify 
the dynamic states of halos. The first is the $\chi^2_{c}$ parameter, 
defined as the minimized $\chi^2$ in fitting the NFW profile, 
normalized by halo mass (see~\S\ref{ssec_concentration_def}). 
The second is the offset parameter, $x_{\rm off}$, defined as the 
distance from the center of mass of the halo to the most bound particle, 
normalized by the virial radius. Figure~\ref{fig:struct_relaxChi2_xoff_dist} 
shows the distributions of $\chi^2_{\rm c}$ 
and $x_{\rm off}$ for subsamples with different last 
major merger times. As one can see, 
if a halo has experienced a recent major merger 
($\log\,\delta_{\rm c}(z_{\rm lmm}) < 0.3$), it is 
likely that its profile deviates from the NFW profile, 
and that its most bound particle is far away from the center of mass of the halo.  

Including those unrelaxed halos in our sample 
will significantly increase the variance of halo properties, 
thereby affecting the statistics derived from the sample.
To reduce their effects, we exclude all halos that have undergone 
a major merger at $z<0.3$. This will   
remove $9.6\%$, $9.6\%$, $14.2\%$, $21.8\%$, $10.5\%$, and $13.8\%$ of 
halos in samples $\rm S_1$, $\rm S_2$, $\rm S_3$, $\rm S_4$,
$\rm S_c$, and $\rm S_L$, respectively. To make our conclusion even less dependent 
on relaxation processes, we compute $\lambda_{\rm s}$ and $q_{\rm axis}$
using only simulated particles that are within a radius $2.5\,r_{\rm s}$
from the most bound particle, where $r_{\rm s}$ is the scale radius of the fitted NFW profile. According to our test, our results 
are not sensitive to the radius chosen.

\bibliography{references}
\bibliographystyle{aasjournal}

\end{document}